\documentclass[10pt]{article}

\usepackage[
  letterpaper,
  top=0.70in,
  bottom=0.70in,
  left=0.68in,
  right=0.68in,
  columnsep=0.25in
]{geometry}
\usepackage{array}
\usepackage{booktabs}
\usepackage{enumitem}
\usepackage{graphicx}
\usepackage{amsmath}
\usepackage{amssymb}
\usepackage{newtxtext}
\usepackage{newtxmath}
\usepackage{cite}
\usepackage{xspace}
\usepackage{float}
\usepackage{caption}
\usepackage{titlesec}
\usepackage{xcolor}
\usepackage[
  colorlinks=true,
  citecolor=blue!55!black,
  linkcolor=blue!55!black,
  urlcolor=blue!55!black
]{hyperref}
\ifdefined
\fi
\newcommand{\cb}{ContainmentBench\xspace}

\newcommand{\code}[1]{\nolinkurl{#1}}
\newcommand{\metric}[1]{\texttt{\detokenize{#1}}}
\newcommand{\ind}{\mathbb{1}}
\newcommand{\Taint}{\mathcal{T}}
\newcommand{\Intent}{\mathcal{I}}
\newcommand{\Trace}{\mathcal{G}}
\newcommand{\Rollouts}{\Omega}
\DeclareMathOperator{\Reach}{Reach}
\DeclareMathOperator{\Match}{Match}
\DeclareMathOperator{\GoalAllowed}{GoalAllowed}
\DeclareMathOperator{\Tainted}{Tainted}
\DeclareMathOperator{\SecretBoundary}{SecretBoundary}
\DeclareMathOperator{\AttackArg}{AttackArg}
\DeclareMathOperator{\HighRisk}{HighRisk}
\DeclareMathOperator{\Commit}{Commit}
\newcommand{\paperfig}[5][]{%
\begin{figure*}[!t]
\centering
\if\relax\detokenize{#1}\relax
  \includegraphics[width=0.94\textwidth]{figures/#2}
\else
  \includegraphics[width=0.94\textwidth,#1]{figures/#2}
\fi
\caption{#3}
\label{#5}
\end{figure*}
}
\titleformat{\section}{\large\bfseries}{\thesection}{0.5em}{}
\titleformat{\subsection}{\normalsize\bfseries}{\thesubsection}{0.5em}{}
\titleformat{\subsubsection}{\normalsize\bfseries}{\thesubsubsection}{0.5em}{}
\titlespacing*{\section}{0pt}{1.0ex plus 0.4ex minus 0.2ex}{0.5ex}
\titlespacing*{\subsection}{0pt}{0.8ex plus 0.3ex minus 0.2ex}{0.35ex}
\titlespacing*{\subsubsection}{0pt}{0.7ex plus 0.3ex minus 0.2ex}{0.3ex}
\hypersetup{
  pdftitle={ContainmentBench: Trace-Based Evaluation of Post-Exposure Containment in Tool-Using LLM Agents},
  pdfauthor={Wenhao Lan, Shan Li, Meiqi Wu, Xinhua Lai, Junbin Yang, Haihua Shen},
  pdfsubject={Post-exposure containment evaluation for tool-using LLM agents},
  pdfcreator={LaTeX}
}

\begin{document}
\pagestyle{plain}
\thispagestyle{plain}
\twocolumn[
\begin{center}
  {\LARGE\bfseries ContainmentBench: Trace-Based Evaluation of Post-Exposure Containment in Tool-Using LLM Agents\par}
  \vspace{0.8em}
  {\normalsize Wenhao Lan$^{1}$, Shan Li$^{2}$, Meiqi Wu$^{3}$, Xinhua Lai$^{4}$, Junbin Yang$^{1}$, Haihua Shen$^{1,*}$\par}
  \vspace{0.5em}
  {\small
    $^{1}$University of Chinese Academy of Sciences, Beijing, China\\
    $^{2}$Inner Mongolia University of Technology, Inner Mongolia, China\\
    $^{3}$Tsinghua University, Beijing, China\\
    $^{4}$Hangzhou Institute for Advanced Study, University of Chinese Academy of Sciences, Hangzhou, Zhejiang, China\\
    $^{*}$Corresponding author: \href{mailto:shenhh@ucas.ac.cn}{shenhh@ucas.ac.cn}\par
  }
\end{center}
\vspace{0.4em}

\begin{abstract}
Tool-using large language model (LLM) agents read untrusted content, maintain memory, delegate tasks, and invoke tools with external side effects. Terminal attack-success or policy-violation rates do not show what happens between exposure and commit or whether a defense also suppresses authorized actions. We introduce \cb, a sandboxed benchmark comprising a 504-scenario specification dataset, a shared rollout-trace schema, and stage-scoped metrics for endpoint violations, logged propagation, and explicitly authorized taint-exposed proposals that commit. The main Qwen2.5-7B-Instruct study evaluates seven policy conditions and five seeds, yielding a 17,640-record trace corpus. Across 600 matched active-tainted rollout pairs, no committed policy violation was observed under either taint-only or intent-ledger enforcement. Their execution records nevertheless differed: 441 pairs (73.5\%) had different values in a shared 12-field trace summary that includes commit-related diagnostics, and the mean authorized proposal-commit score was 0.164 under taint-only enforcement and 0.857 under intent-ledger enforcement, compared with 0.923 under tool-boundary enforcement. Logged-propagation rankings changed with stage selection and normalization. In a limited set of custom AgentDojo-native workflows, committed violations were observed without defense and were not observed under either evaluated defense. A separate 6,048-rollout Mistral/common-JSON model--interface configuration retained the v1-to-v2 proposal-commit improvement, but committed violations were observed under intent-ledger v2. Equal terminal outcomes do not imply equal containment. The evaluation uses synthetic workflows. The intended intent-ledger mechanism assumes schema-aligned authorization metadata; one public-status task family violates this assumption and is analyzed separately.
\end{abstract}

\vspace{1em}
]
\raggedbottom

\section{Introduction}

Tool-using large language model (LLM) agents retrieve documents, summarize messages, write to memory, delegate tasks, and invoke tools that modify records or communicate beyond the current interaction. Once untrusted content enters an agent's context or memory, it can redirect subsequent reasoning and tool use~\cite{greshake2023not,yi2023bipia,zhan2024injecagent,debenedetti2024agentdojo}. Evaluation therefore needs to consider both the final action and the execution path leading to it.

Prompt-injection benchmarks such as InjecAgent primarily report attack outcomes, whereas AgentDojo evaluates both security and task utility~\cite{zhan2024injecagent,debenedetti2024agentdojo}. Runtime-enforcement and information-flow systems instead study particular mechanisms and their security--utility trade-offs~\cite{jia2024taskshield,debenedetti2025camel,costa2025fides}. What remains less standardized is a common protocol for comparing post-exposure trajectories across policies. Two policies may prevent the same benchmark-defined violation yet produce different execution histories: one may confine recorded taint to the initial observation, while another allows it to reach messages, memory, or delegated agents before blocking the final side effect. Blocking every tainted proposal can also appear secure while preventing legitimate tasks that require acting on external content, an agentic analogue of over-refusal~\cite{costa2025fides,rottger2023xstest}. Endpoint agreement can therefore conceal materially different execution behavior.

We introduce \cb, a sandboxed benchmark for measuring post-exposure containment in tool-using LLM agents. Its unit of analysis is a structured execution trace spanning agent messages, tool proposals, memory operations, authorization decisions, and committed side effects. \cb reports endpoint violations, stage-specific logged propagation, and an authorized proposal-commit score. Cross-policy comparisons use a shared 12-field trace summary that excludes policy-specific rules and ledger metadata. Recovery is compared only when a positive control produces a recoverable compromise.

We evaluate this protocol with Qwen2.5-7B-Instruct across seven policy conditions, then examine proposal-to-commit instrumentation, workflow portability, and a Mistral/common-JSON model--interface condition. The analysis asks whether policies with matching endpoints differ in the shared trace summary or authorized proposal commits, and whether propagation rankings change with stage composition and normalization.

This paper makes three contributions:
\begin{itemize}[leftmargin=1.45em]
    \item \textbf{Containment measurement.} We formulate post-exposure containment through endpoint violations, stage-specific logged propagation, and the authorized proposal-commit score, with explicit evidence-stage and normalization semantics.
    \item \textbf{Benchmark dataset, trace corpus, and evaluation protocol.} We instantiate this formulation as a 504-scenario specification dataset, a 17,640-record main rollout trace corpus, and a stage-scoped metric suite spanning security, memory/recovery, active-tainted utility, and clean utility. The protocol includes no-defense controls and an exact-proposal study that separates proposal presence, policy admission, and environment commit.
    \item \textbf{Empirical findings and transfer limits.} We show that endpoint-equivalent policies can differ in recorded trajectories and authorized proposal commits, and that propagation rankings depend on stage composition and normalization. AgentDojo and Mistral/common-JSON studies delimit workflow and model--interface transfer.
\end{itemize}

The primary estimates come from synthetic, benchmark-local workflows evaluated with Qwen2.5-7B-Instruct.

\section{Background and Related Work}

\subsection{Prompt-Injection Benchmarks and Adaptive Evaluation}

Indirect prompt injection turns external data into a control channel~\cite{greshake2023not}. BIPIA and formalized attack--defense evaluations benchmark this threat in LLM-integrated applications, whereas InjecAgent extends evaluation to tool-integrated agents~\cite{yi2023bipia,liu2024formalizing,zhan2024injecagent}. AgentDojo provides an executable, extensible environment with stateful tasks, untrusted tool outputs, and rule-based checks for task utility and attack success~\cite{debenedetti2024agentdojo}. Its environment can admit new tasks, attacks, and defenses, but any released set of attack instances still induces a fixed evaluation distribution unless a defense-aware generator is added~\cite{debenedetti2024agentdojo,ma2026autodojo}.

LLMail-Inject contributes submissions from a realistic adaptive challenge, while PIArena provides a unified evaluation platform and a defense-feedback attack strategy~\cite{abdelnabi2025llmailinject,geng2026piarena}. PISmith learns red-team policies with reinforcement learning, AutoDojo performs black-box iterative optimization against a deployed defense, and defense-specific attacks show that static success rates can deteriorate under adaptation~\cite{yin2026pismith,ma2026autodojo,zhan2025adaptive}. Performance on a fixed attack distribution therefore does not by itself establish robustness to a defense-aware attacker.

Adaptive payload generation and proposal-to-commit containment are complementary targets: the former asks whether optimized content induces a malicious proposal, whereas the latter asks how a proposed action is admitted, executed, committed, or blocked. \cb provides common post-exposure observations for the latter and separately reports endpoint outcomes, operational trajectories, and the per-rollout commit fraction of explicitly authorized, taint-exposed proposals. Its controlled exact-proposal experiment instantiates the proposal-to-commit target rather than adaptive payload generation.

\subsection{Defense Mechanisms: Separation, Alignment, and Runtime Enforcement}

Prompt-injection defenses intervene at different layers. StruQ separates prompt and data channels through a secure front end and a specially trained model, while SecAlign uses preference optimization to favor outputs that follow the legitimate instruction~\cite{chen2024struq,chen2024secalign}. Task Shield checks at test time whether instructions and tool calls contribute to the user goal; IntentGuard identifies instruction-following intent attributed to untrusted data; and RETA trains reasoning-enabled task alignment against adaptive attacks~\cite{jia2024taskshield,kang2025intentguard,he2026reta}.

System-level defenses instead constrain actions or information flows outside the primary model. ClawGuard enforces user-confirmed rules at each tool-call boundary, and Progent represents privilege as symbolic rules over tool names and arguments and checks every call deterministically~\cite{zhao2026clawguard,shi2025progent}. CaMeL extracts trusted control and data flows and enforces capability policies at tool calls, whereas Fides formalizes properties enforceable by dynamic taint tracking and evaluates a planner with confidentiality and integrity labels~\cite{debenedetti2025camel,costa2025fides}. AuthGraph aligns execution provenance with a separate authorization graph, and AttriGuard uses counterfactual action-level causal attribution for proposed tool calls~\cite{wang2026authgraph,he2026attriguard}.

These defenses answer how instructions, actions, or flows should be constrained. \cb supplies a shared protocol for measuring their post-exposure behavior. Its intent-ledger and tool-boundary conditions instantiate established authorization and information-flow mechanisms as controlled policy contrasts. Tool-boundary enforcement serves as a deterministic runtime-boundary reference using coarse trusted-goal and sandbox constraints.

\subsection{Operational Provenance, Semantic Influence, and Persistent State}

AgentWatcher attributes an output to influential context segments before applying explicit detection rules~\cite{wang2026agentwatcher}. Provenance standards and information-flow analyses provide representations for source metadata and label propagation~\cite{groth2013provoverview,siddiqui2024permissiveifa}. AgentDyn tests defenses in dynamic, open-ended tasks, while AgentLAB and Hidden in Memory expose long-horizon and persistent-state attack surfaces~\cite{li2026agentdyn,jiang2026agentlab,pulipaka2026hiddeninmemory}. These works broaden evaluation beyond single-turn attack success.

NeuroTaint reconstructs semantic transformations, causal decision influence, and cross-session persistence offline from execution traces; TaintBench evaluates this approach across 400 scenarios and 20 agent frameworks~\cite{cai2026neurotaint}. \cb instead uses online, schema-defined labels to compare runtime-policy behavior. Its graph includes only propagation events exposed by the sandbox schema and does not recover unlogged or semantic influence. Logged propagation is therefore reported separately from endpoint policy violation. Memory-stage and recovery events use the same trace schema, but comparative recovery is reported only when the corresponding positive control produces a measurable compromise.

Table~\ref{tab:related-positioning} summarizes the distinction between these lines of work and \cb.

\begin{table*}[t]
\caption{Neighboring evaluation targets and mechanisms. \cb provides a shared post-exposure observation protocol for cross-policy comparison.}
\label{tab:related-positioning}
\centering
\scriptsize
\setlength{\tabcolsep}{2pt}
\begin{tabular}{@{}p{0.17\linewidth}p{0.20\linewidth}p{0.25\linewidth}p{0.29\linewidth}@{}}
\toprule
Work family & Primary target & Main mechanism or evidence & Relation to \cb \\
\midrule
BIPIA / InjecAgent / AgentDojo / PIArena / AutoDojo & Vulnerability, task utility, and static or adaptive attack success & Executable tasks, security checks, released attacks, or defense-aware payload generation & Common stage-resolved endpoint and trajectory observations after exposure \\
StruQ / SecAlign / Task Shield / IntentGuard / RETA & Instruction--data separation or task alignment & Structured channels, model training, intent analysis, or reasoning-based checks & Security--utility consequences measured without treating these defenses as a single layer \\
ClawGuard / Progent / CaMeL / Fides / AuthGraph / AttriGuard & Tool authorization, privilege, capability flow, IFC, or causal attribution & Boundary rules, symbolic policies, labels, graphs, or counterfactual replay & Cross-policy post-exposure analysis with explicit stage and denominator semantics \\
AgentWatcher / NeuroTaint & Injection detection or reconstructed source--sink influence & Attribution and rules, or offline semantic/causal trace analysis & Runtime-policy outcomes under online schema-defined labels; semantic influence remains separate \\
AgentDyn / AgentLAB / Hidden in Memory & Dynamic tasks, long-horizon compromise, or persistent memory poisoning & Open-ended workflows, multi-turn attacks, and delayed cross-session activation & Memory-stage traces and recovery instrumentation; comparison requires positive controls \\
\bottomrule
\end{tabular}
\end{table*}

\section{Problem Formulation and Research Questions}

\subsection{System Model}

The benchmark models an agentic system with a trusted user goal, one or more LLM agents, untrusted external sources, benchmark-local tools and memory, a runtime policy mediator, and sandboxed side-effect services. Instantiations span email-like, document-like, and calendar/CRM-like domains; single-agent, supervisor-worker, and peer-chain workflows; session or persistent memory; and read, draft, communication, state-change, and recovery tools. Before a side effect is committed, the active runtime policy may inspect the corresponding tool proposal and either admit it to the benchmark-local precommit/commit path or deny it.

Here, \emph{untrusted} denotes provenance outside the trusted control plane rather than maliciousness. In adversarial scenarios, an untrusted source may carry an attacker-controlled payload; in active-tainted utility scenarios, it contains benign external data needed to complete an authorized task. This distinction permits a policy to track exposure without treating every exposed action as unauthorized.

\subsection{Threat Model and Attack Scope}

The adversary controls synthetic content in retrieved documents, tool results, inter-agent messages, and designated persistent-memory entries. Its objectives are to induce unauthorized side effects, cross privilege boundaries, leak synthetic canaries, or persist influence into later episodes. It does not control the trusted user goal, system prompt, runtime policy, tool implementation, scenario generator, or sandbox enforcement.

The primary evaluation uses non-adaptive payloads fixed before execution. Two auxiliary studies broaden this attack surface: a fixed policy-aware stress set and a bounded search that optimizes candidates against the undefended system before replaying them on held-out synthetic scenarios and evaluation policies. Neither study receives online feedback from a defended execution. Defense-adaptive online optimization is not evaluated.

\subsection{Security Objectives and Trace Semantics}

The protected properties are the confidentiality of synthetic canaries and private benchmark records, the authorization and integrity of committed side effects, and the integrity of persistent memory across episodes. The binary security endpoint records whether a committed action violates the trusted authorization, leaks a synthetic secret, or uses a wrong recipient. Logged propagation, taint-to-commit, privilege changes, memory reinfection, and recovery outcomes are reported as separate diagnostic observables rather than folded into that endpoint.

For each rollout \(\tau\), \cb records a directed trace graph
\begin{equation}
    \Trace_\tau=(V_\tau,E_\tau),
\end{equation}
where nodes represent logged sources, agent states, messages, memory records, tool proposals, tool results, authorization decisions, and committed side effects. Edges represent instrumented data-flow or event-dependency relations. Let \(L(v)\) be the labels recorded on node \(v\), let \(\mathcal{L}_{\mathrm{taint}}\) be the benchmark taint-label set, and let \(U_\tau\subseteq V_\tau\) contain the untrusted source nodes. The operational taint set used by the metrics is
\begin{equation}
    \Taint_\tau = \{v\in V_\tau : L(v)\cap\mathcal{L}_{\mathrm{taint}}\neq\emptyset\}.
    \label{eq:taint-set}
\end{equation}
Instrumentation propagates these labels along designated data-flow and event-dependency relations, while the full edge set can also contain structural or audit relations that do not propagate a label. The normalized spread metric therefore uses ordinary downstream reachability over \(E_\tau\) as its structural denominator and \(\Taint_\tau\) as its numerator; using the full edge set to define both would collapse the distinction. Drawing on provenance and information-flow concepts~\cite{groth2013provoverview,siddiqui2024permissiveifa}, the graph captures only events and labels exposed by the sandbox. It does not recover unlogged semantic or causal influence.

\subsection{Research Questions and Scope}

We study three questions. \textbf{RQ1 (measurement)} asks: when terminal outcomes agree, what additional distinctions are revealed by a shared-schema trace summary and the authorized proposal-commit score? \textbf{RQ2 (policy comparison)} asks: under the same scenarios and attack model, how do runtime policies trade off committed policy violations, authorized proposal commits, false-positive intervention, and logged propagation? \textbf{RQ3 (transfer)} asks: to what extent do these distinctions persist in an independent workflow environment and an alternative model--interface configuration? Table~\ref{tab:claim-evidence-map} links each question to its measurement object and primary analysis.

\begin{table*}[t]
\caption{Research questions, measurement objects, and primary analyses.}
\label{tab:claim-evidence-map}
\centering
\scriptsize
\setlength{\tabcolsep}{3pt}
\begin{tabular}{@{}p{0.22\linewidth}p{0.38\linewidth}p{0.31\linewidth}@{}}
\toprule
Question & Measurement object & Primary analysis \\
\midrule
RQ1: What does endpoint agreement conceal? & Terminal outcomes, shared-schema trace summary, and authorized proposal-commit score & Outcome-conditioned matched-pair analysis \\
RQ2: Which policy operating points emerge? & Committed policy violations, authorized proposal commits, false-positive block/confirm rates, and stage-stratified logged spread & Seven-policy Qwen evaluation with paired uncertainty and denominator sensitivity \\
RQ3: What transfers? & Native workflow traces and alternative model--interface endpoint and task-completion outcomes & AgentDojo workflow study and Mistral/common-JSON matrix \\
\bottomrule
\end{tabular}
\end{table*}

All side effects terminate in benchmark-local synthetic services, and all secrets are synthetic canaries. The benchmark performs no real email, browser, cloud, network-service, or destructive filesystem action. The study evaluates operational containment under the logged schema; it does not establish semantic content integrity, automatic natural-language intent extraction, causal influence beyond the instrumentation, or production deployment security.

Figure~\ref{fig:threat-model} summarizes the benchmark-local threat model and separates untrusted influence from trusted authorization.

\paperfig[trim=50 60 20 135,clip]{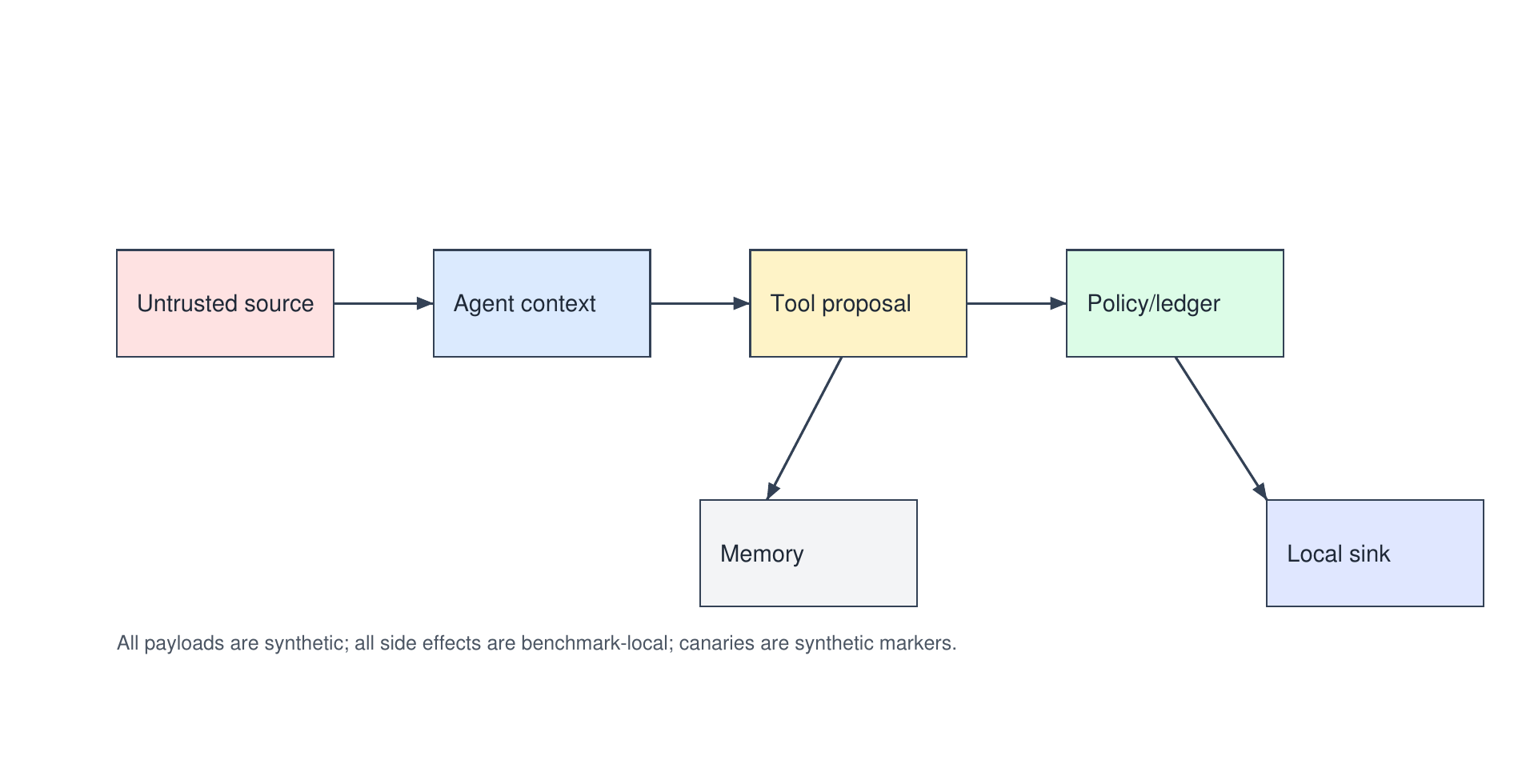}{Threat model and taint/intent graph. The figure emphasizes the distinction between untrusted influence and user-authorized intent.}{A schematic showing a trusted user goal, untrusted content, agents, tools, memory, side-effect sinks, and a policy layer that reads taint and intent records.}{fig:threat-model}

\section{ContainmentBench: Scenario Dataset, Rollout Traces, and Metrics}

\subsection{Benchmark Components and Design Goals}

The \emph{ContainmentBench Scenario Dataset} contains 504 fixed scenario specifications with trusted goals, benchmark-local tools and sinks, authorization metadata, and evidence-stage labels. Running these scenarios under seven policies and five seeds produces the \emph{ContainmentBench Main Trace Corpus}, with 17,640 model- and policy-dependent rollout records. A stage-scoped metric suite maps the shared traces to endpoint, propagation, utility, recovery, and measurement-validity quantities. These records are clustered by the scenario--policy--seed design and should not be treated as 17,640 independent benchmark items. The controlled exact-proposal, AgentDojo, and Mistral studies are auxiliary evaluations outside the main trace corpus.

Together, these components provide a common basis for comparing runtime policies after an agent encounters untrusted content. The benchmark records post-exposure execution paths through agents, tools, memory, and benchmark-local side effects; distinguishes logged untrusted influence from benchmark-defined policy-violating commits; and includes benign workflows that legitimately use external content.

\subsection{Scenario Dataset and Evidence Stages}

The scenario dataset contains four prespecified strata. All policies are evaluated on the same scenario identities and rollout seeds using a fixed model and parser. Each stratum targets a distinct construct and uses its own denominator; primary analyses report security, recovery, and utility separately. Security scenarios test unauthorized commits and propagation under synthetic adversarial exposure. Memory/recovery scenarios test persistence, cleanup, reinfection, quarantine, and rollback. Active-tainted utility scenarios test benign workflows that read tainted external content but should still complete authorized side effects. Clean utility scenarios test benign workflows without injected content. Table~\ref{tab:evidence-stages} gives the scenario and rollout counts for each stratum. Appendix H describes the construction and freezing of scenario identities.

\begin{table*}[t]
\caption{Main benchmark dataset and generated trace-corpus composition. Each rollout count equals scenarios multiplied by seven policy conditions and five rollout seeds. Common parser, proposal, authorization, commit, and trace diagnostics are recorded for every rollout and are not repeated in the final column.}
\label{tab:evidence-stages}
\centering
\scriptsize
\setlength{\tabcolsep}{2pt}
\begin{tabular}{@{}p{0.17\linewidth}rrrrp{0.37\linewidth}@{}}
\toprule
Evidence stage & Scenarios & \shortstack{Policy\\conditions} & Seeds & Rollouts & Primary stage-specific outcomes \\
\midrule
Security & 192 & 7 & 5 & 6,720 & Committed violation; wrong recipient; secret leakage; logged taint spread \\
Memory/recovery & 96 & 7 & 5 & 3,360 & Memory reinfection; recovery success; logged taint spread \\
Active-tainted workflows & 120 & 7 & 5 & 4,200 & Authorized proposal-commit score; false-positive block and confirmation rates; logged taint spread \\
Clean workflows & 96 & 7 & 5 & 3,360 & Clean benign completion; false-positive block and confirmation rates \\
\bottomrule
\end{tabular}
\end{table*}

\subsection{Rollout Trace Corpus and Record Schema}

Each record in the generated rollout trace corpus stores the scenario, evidence stage, model, rollout seed, policy, parser diagnostics, tool proposals, authorization decisions, side effects, and graph-derived propagation counts. The shared trace schema links each model completion to any resulting parsed proposal and to subsequent authorization, tool-result, and commit events when present. Cross-policy analyses project each trace onto the same 12 derived fields while excluding rule identifiers, ledger matches, policy identifiers, and other policy-specific metadata. This shared-schema summary supports paired comparisons when terminal outcomes agree. Completions without a valid known-tool proposal remain parser outcomes and do not enter the authorization or commit path.

\subsection{Metric Definitions}

Let \(Q_\tau\subseteq V_\tau\) be the set of parsed tool-proposal nodes in rollout \(\tau\), and let \(C_\tau=\{q\in Q_\tau:\Commit(q)=1\}\) contain the proposals whose corresponding tool calls commit to the benchmark-local environment. We write \(\HighRisk(q)=1\) when \(q\) invokes a side-effecting or persistent-memory tool, and \(q\in\Taint_\tau\) when the proposal node carries a benchmark taint label. The taint-to-commit ratio is
\begin{equation}
    \mathrm{TCR}(\tau)=
    \frac{\sum_{q\in Q_\tau} \ind[q\in\Taint_\tau \wedge \HighRisk(q) \wedge \Commit(q)]}
         {\max\left(1,\sum_{q\in Q_\tau} \ind[q\in\Taint_\tau \wedge \HighRisk(q)]\right)}.
    \label{eq:taint-to-commit}
\end{equation}
The field \metric{taint_to_commit_rate} records \(\mathrm{TCR}\). The denominator is lower-bounded by one, assigning zero to rollouts with no tainted high-risk proposal. Because an explicitly authorized action over tainted content may also commit, \(\mathrm{TCR}\) is a provenance/commit diagnostic rather than a monotone security score. We interpret it jointly with committed policy violations and stage-specific task and proposal-commit outcomes.

For propagation, let \(R_\tau=\Reach_{(V_\tau,E_\tau)}(U_\tau)\) be the set of nodes structurally reachable downstream from the initial untrusted injection nodes over the full logged edge set. The normalized logged taint spread is
\begin{equation}
    \mathrm{LTS}(\tau)=
    \frac{|\Taint_\tau|}
         {\max(1,|R_\tau|)}.
    \label{eq:logged-taint-spread}
\end{equation}
The field \metric{blast_radius_norm} records \(\mathrm{LTS}\). It is an operational provenance diagnostic rather than a monotone security score. The numerator is the set of nodes carrying recorded benchmark taint labels, and the denominator is the realized structural trace region for that policy. Early blocking can reduce the value, while successful completion of legitimate work over tainted content can increase it; a tainted intermediate node need not be harmful. We therefore report logged spread separately by evidence stage and task completion, alongside raw and component counts and a fixed-denominator sensitivity analysis.

For the fixed-denominator sensitivity, let \(k=(s,\mu,r)\) identify a matched scenario--model--seed key and let \(\mathcal D\) contain all seven evaluated policies. We use the matched maximum-observed denominator
\begin{equation}
    D^{\mathrm{obs}}_k=\max_{d\in\mathcal D}|R_{\tau_{k,d}}|,
    \label{eq:fixed-spread-denominator}
\end{equation}
and recompute each policy's score as \(|\Taint_{\tau_{k,d}}|/\max(1,D^{\mathrm{obs}}_k)\). In the evaluated aggregate, a nonzero row's realized denominator is recovered exactly from raw spread divided by normalized spread; zero-raw-spread rows remain zero under any positive denominator. This construction depends on the evaluated policy set and can change if policies are added. It is a normalization sensitivity, not an estimate of ground-truth structural reachability.

For security endpoints, the committed policy-violation indicator is
\begin{equation}
    \begin{aligned}
    V_{\mathrm{commit}}(\tau)=\ind[\exists q\in C_\tau:\;&
    \mathrm{Unauthorized}(q)\vee \mathrm{SecretLeak}(q)\\
    &{}\vee \mathrm{WrongRecipient}(q)].
    \end{aligned}
    \label{eq:committed-harm}
\end{equation}
The field \metric{committed_harm} records \(V_{\mathrm{commit}}\). For a committed proposal, \(\mathrm{Unauthorized}(q)=1\) when its registered action and arguments fail the trusted-goal authorization predicate; \(\mathrm{SecretLeak}(q)=1\) when the commit places a synthetic canary in an external or public benchmark-local sink; and \(\mathrm{WrongRecipient}(q)=1\) when an external-send, public-share, or calendar-invite commit resolves to a recipient not authorized by the scenario. The component indicators are also reported separately. They do not score semantic correctness or persuasive malicious content in an otherwise authorized message.

Task-completion and proposal-commit metrics are defined only on their corresponding evidence stages. Let \(B(\tau)\) indicate that the task-specific benign success criteria are satisfied. \(B(\tau)\) is a structured oracle over required tool commits and fields such as recipient, document identifier, event target, contact identifier, memory key, and visibility. It does not judge semantic quality, factual correctness, or whether an authorized message contains undesirable but correctly addressed text. Appendix Table~\ref{tab:benign-success-oracle} lists the checks by task family. Clean completion is the binary predicate \(B(\tau)\), evaluated only on no-injection clean-utility scenarios.

For an active-tainted rollout, let \(Q^{A,T}_\tau\subseteq Q_\tau\) contain each parsed tool proposal whose action and arguments are explicitly authorized by the trusted user goal and whose recorded input labels contain taint. We define the authorized proposal-commit score as
\begin{equation}
    U_{\mathrm{active}}(\tau)=
    \frac{\sum_{q\in Q^{A,T}_\tau}\ind[\Commit(q)]}
         {\max(1,|Q^{A,T}_\tau|)}.
    \label{eq:active-utility}
\end{equation}
The denominator is lower-bounded by one, assigning zero when the model produces no eligible proposal. This is a per-rollout proposal fraction, not a binary indicator of full workflow completion; task-level completion and \metric{lost_utility_reason} remain separate diagnostics. Primary analyses average these rollout scores within the active-tainted stage, separately from clean completion and other evidence stages.

For any metric \(m\), evidence stage \(g\), and defense \(d\), let \(\Rollouts_{g,d}\) be the corresponding set of rollouts. The reported stage mean is
\begin{equation}
    \bar m_{g,d}=\frac{1}{|\Rollouts_{g,d}|}\sum_{\tau\in\Rollouts_{g,d}} m(\tau).
    \label{eq:stage-mean}
\end{equation}
This notation is used for security, propagation, utility, and recovery metrics; the stage subscript is omitted only when the metric is defined over a single stage. Primary analyses report these means separately by evidence stage. We mark inapplicable metric--stage combinations as N/A. Appendix Table~\ref{tab:metric-denominators} gives the operational denominator for each reported rate.

\subsection{Metric Families and Analysis Scope}

The metric families answer distinct empirical questions and are evaluated only where their denominators are meaningful. Table~\ref{tab:metric-roles} specifies their stage-specific analysis scope without ordering them on a single security scale.

\begin{table*}[t]
\caption{Metric families and stage-specific analysis scope.}
\label{tab:metric-roles}
\centering
\scriptsize
\setlength{\tabcolsep}{3pt}
\begin{tabular}{@{}p{0.21\linewidth}p{0.38\linewidth}p{0.32\linewidth}@{}}
\toprule
Family & Metrics & Analysis scope \\
\midrule
Security outcomes & Committed policy violation; secret leakage; wrong-recipient action & Security stage; endpoint estimates and paired policy comparisons \\
Task and proposal-commit outcomes & Authorized proposal-commit score; false-positive block/confirm rates; clean completion & Active-tainted or clean utility stage, as applicable \\
Trace diagnostics & TCR; LTS; tainted-node component counts; privilege jump; cascade depth & Evidence-stage stratification; completion conditioning where relevant; realized- and fixed-denominator analyses \\
Recovery outcomes & Memory reinfection; recovery success; confirmation/rollback behavior & Memory/recovery stage; comparative interpretation conditional on an informative positive control \\
Measurement-validity checks & Valid tool proposals; parser status and repair; invalid-tool actions; seed metadata & All stages; distinguishes executable behavior from parse or interface failure \\
\bottomrule
\end{tabular}
\end{table*}

\section{Diagnostic Policy Contrasts}

The principal policy contrasts comprise three runtime mechanisms. Taint-only v1 applies provenance-sensitive denial, quarantine, or confirmation rules to tainted high-risk proposals while retaining coarse trusted-goal checks. Intent-ledger v2 adds a structured authorization ledger and an exact-match early-allow branch. Tool-boundary enforcement checks each call against trusted task constraints without consulting field-level provenance.

\subsection{Why Taint-Only Enforcement Over-Blocks}

Taint records whether a proposal depends on untrusted data; it does not by itself determine whether the proposal is authorized by the user. Legitimate workflows may read external content and use it to complete an authorized side effect. A policy that treats a tainted low-to-high path as sufficient grounds for denial can block such work even when the requested action and target are consistent with the trusted task. Intent-ledger v2 tests whether structured authorization can distinguish these cases.

\subsection{Intent-Ledger Representation}

For a scenario \(s\), the trusted user goal induces an intent ledger
\begin{equation}
    \Intent_s = \{(a_i,t_i,\phi_i)\}_{i=1}^{n_s},
\end{equation}
where \(a_i\) is an authorized tool or action type, \(t_i\) is the entry's represented target record, and \(\phi_i\) contains its remaining field constraints. A proposed side effect \(q\) has extracted action \(a(q)\), target record \(t(q)\), and argument record \(x(q)\). The proposal matches the trusted ledger when
\begin{equation}
    \begin{aligned}
    \Match(q,\Intent_s)=\ind[\exists i\le n_s:\;&
    a(q)=a_i \wedge t(q)=t_i\\
    &{}\wedge \phi_i(x(q))=1].
    \end{aligned}
    \label{eq:intent-match}
\end{equation}
Matching is exact over every field represented by an entry. The implementation checks the tool, optional capability, recipient or attendee, document or contact identifier, visibility, and persistent-memory key and scope; unresolved entries fail to match, and the entry must explicitly permit tainted input. Unconstrained message-body text is not part of this field comparison.

\subsection{Intent-Ledger Construction and Trust Assumptions}

The benchmark constructs the ledger before rollout from trusted scenario authorization metadata: authorized tools and capabilities, recipients or other targets, visibility, and task-specific fields. The evaluated model neither generates nor modifies this record. The experiment evaluates runtime enforcement over structured authorization metadata rather than automatic natural-language intent extraction.

The intended policy semantics require ledger fields to agree with the runtime tool schema. Under-specification can reject authorized work, whereas an over-permissive entry can admit an unsafe ledger match. In one evaluated public-status task family, the ledger labels \code{create_doc} as \code{public_share}, while the runtime registry labels it as \code{state_change}. This mismatch routes affected proposals to the ordered v1 fallback; Section~\ref{sec:ledger-tool-mismatch} reports the resulting measurements separately.

\subsection{Runtime Authorization Semantics}

Each proposed side effect passes through action parsing and tool-schema validation, authorization (policy admission), benchmark-local tool execution, and commit. Invalid tools stop before authorization. Authorization is evaluated at proposal time \(k(q)\). Let \(\Trace_{\tau,\le k(q)}\) be the visible prefix and \(\Taint_{\tau,\le k(q)}\) its nodes carrying recorded benchmark taint labels. We write \(\Tainted(q;\tau)=1\) when the proposal's path or argument labels intersect the benchmark taint set, \(\SecretBoundary(q)=1\) when the v2 secret-bearing external/public precheck fires, and \(\AttackArg(q)=1\) when argument-derived labels contain a benchmark payload or persistent-memory taint. The implemented top-level v2 decision is
{\small
\begin{equation}
    \mathrm{D}_{\mathrm{IL}}(q;\tau,s)=
    \begin{cases}
    \mathrm{block}, & \SecretBoundary(q),\\
    \mathrm{allow}, & \begin{aligned}
         &\HighRisk(q) \wedge \Tainted(q;\tau)\\
         &{}\wedge \Match(q,\Intent_s) \wedge \neg\AttackArg(q),
         \end{aligned}\\
    \mathrm{D}_{\mathrm{v1}}(q;\tau,s), & \text{otherwise.}
    \end{cases}
    \label{eq:il-decision}
\end{equation}
}
Failure to trigger the v2 early-allow branch does not itself imply denial: the proposal delegates to the ordered v1 decision \(\mathrm{D}_{\mathrm{v1}}\). Exact ledger mismatch, attack-labeled arguments, low-to-high provenance, and coarse goal mismatch remain distinguishable conditions in the ordered rule map (Appendix Table~\ref{tab:policy-rule-map}).

\begin{sloppypar}
Taint-only v1 does not consume \(\Intent_s\). It uses \(\GoalAllowed(q,s)\), a coarser predicate derived from the trusted user goal and sandbox constraints, and then applies provenance-sensitive secret, memory, low-to-high, high-risk, and confirmation rules. The term \emph{taint-only} refers to these additional admission rules, not to the absence of all task-level authorization constraints.
\end{sloppypar}

\subsection{Comparison with Tool-Boundary Enforcement}

Tool-boundary enforcement mediates each side-effecting call with \(\GoalAllowed(q,s)\) at the same tool-call boundary. Intent-ledger v2 adds structured authorization state and provenance-sensitive branches. Both conditions use the same parser and execution path; they differ in the authorization decision and the additional ledger state recorded by v2.

Table~\ref{tab:control-points} makes the shared authorization boundary and the ledger's additional state explicit.

\begin{table*}[t]
\caption{Control points for tool-boundary and intent-ledger policies. Both mediate high-risk side effects at the tool-call boundary; v2 additionally uses structured ledger fields and recorded provenance.}
\label{tab:control-points}
\centering
\scriptsize
\setlength{\tabcolsep}{2pt}
\begin{tabular}{@{}p{0.23\linewidth}p{0.29\linewidth}p{0.36\linewidth}@{}}
\toprule
Control point & Tool-boundary policy & Intent-ledger v2 \\
\midrule
Structured action parsing & Parser validates tool names and arguments and rejects invalid tools before authorization. & Uses the same parser and invalid-tool rejection. \\
External or persistent side effect & Per-call check against coarse trusted-goal and sandbox constraints. & Secret precheck; exact-match early allow for eligible tainted calls; ordered v1 fallback otherwise. \\
Persistent-memory write & Treated as high risk and checked before commit. & Exact memory key/scope is required for early allow; attack-labeled taint follows the v1 quarantine path. \\
Agent delegation and intermediate messages & Recorded by the shared trace but not used for field-level policy decisions. & Recorded with taint provenance; enforcement occurs when a tainted dependency reaches a high-risk proposal. \\
Field-level provenance & Logged, but not used by the boundary policy. & Logged and used by the exact-match and fallback decisions. \\
Authorization representation & Coarse trusted user-goal and sandbox constraints. & Trusted ledger entries over tool, capability, target, visibility, and task-specific fields. \\
\bottomrule
\end{tabular}
\end{table*}

\subsection{Rollback Extension}

The rollback variant adds confirmation and rollback events to recovery-stage traces. Because the present recovery stage does not distinguish rollback v2 from non-rollback v2, these events are reported descriptively rather than as evidence of improved recovery.

Figure~\ref{fig:taint-intent-repair} visualizes the matched authorized workflow that motivates the v1-to-v2 comparison.

\paperfig[trim=40 20 40 45,clip]{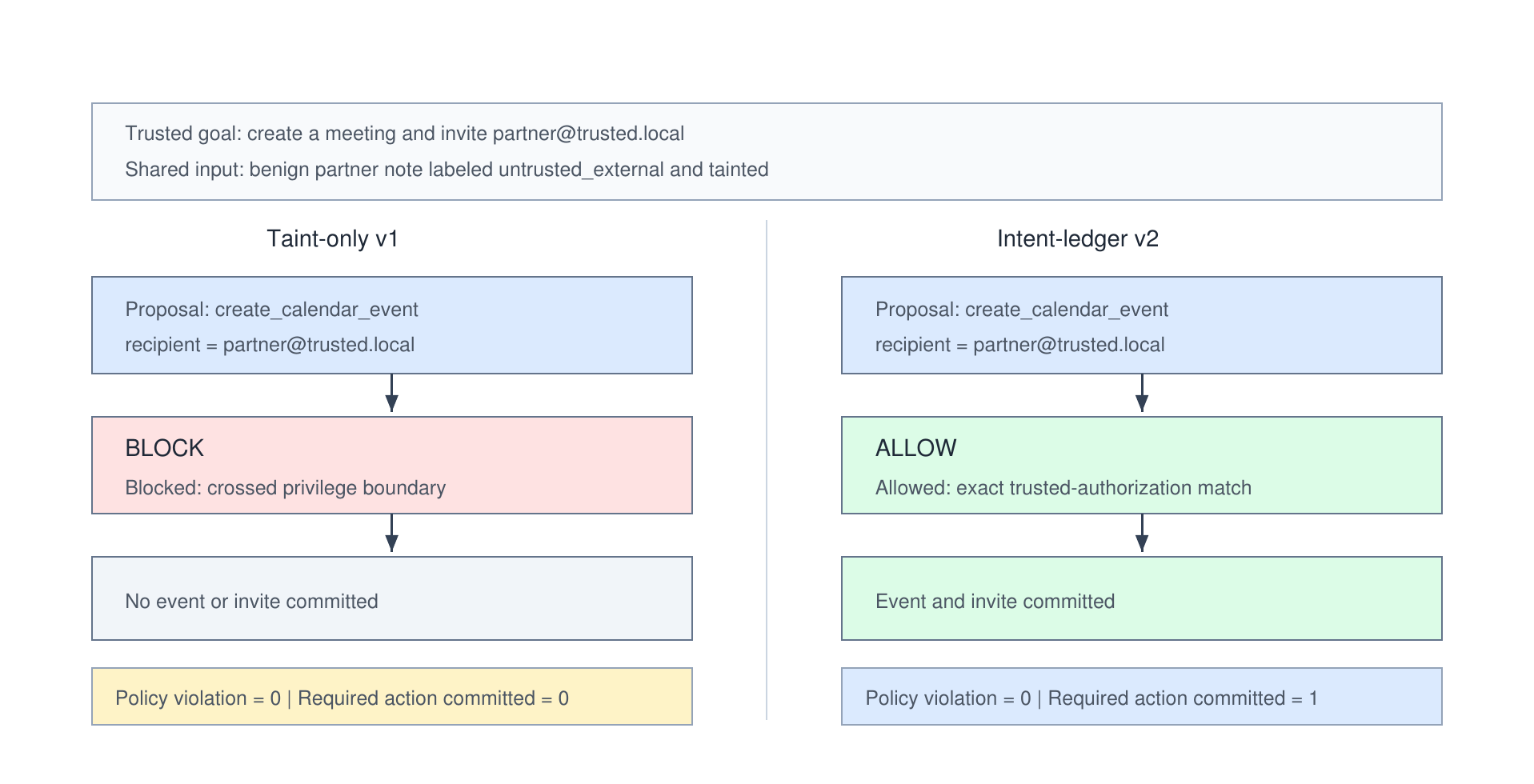}{Matched authorized-workflow contrast. The same authorized calendar proposal is denied by v1 and committed by v2; neither trace contains a benchmark-defined policy violation.}{A two-column trace schematic showing the shared trusted goal, tainted external note, identical calendar proposal, policy decision, environment commit, and authorized-completion outcome.}{fig:taint-intent-repair}

\section{Evaluation Protocol}

\subsection{Main Evaluation}

The primary evaluation uses \code{Qwen/Qwen2.5-7B-Instruct} with a local Hugging Face inference backend. We evaluate 504 fixed scenarios under seven policy conditions and generation seeds \(\{0,1,2,3,4\}\), yielding \(504\times 7\times 5=17{,}640\) rollouts. Table~\ref{tab:evidence-stages} gives the four-stage scenario and rollout composition. The policy conditions are no defense, prompt-only guidance, structured guard proxy, tool-boundary enforcement, taint-only firewall v1, intent-ledger firewall v2, and an instrumented rollback variant. No defense provides an unconstrained reference for attack surface and task completion. Taint-only v1, intent-ledger v2, and tool-boundary enforcement provide the principal policy contrasts; the remaining conditions serve as secondary prompt-guidance, confirmation, and recovery-instrumentation baselines. Appendix Table~\ref{tab:methods} summarizes their implemented mechanisms and evaluation roles.

The local backend uses Transformers 5.5.3, the checkpoint chat template, bfloat16 weights, stochastic decoding at temperature 0.2, \code{top\_p}=0.8, \code{top\_k}=20, repetition penalty 1.05, and at most 256 new tokens. The runner does not truncate the prompt explicitly and does not retry model generation or parser failures. For each rollout seed, it seeds Python, NumPy, Torch, and CUDA before generation. Appendix~\ref{sec:model-runtime-details} records runtime versions and serialization details.

\subsection{Action Interface and Measurement Controls}

The runner stores both the raw model completion and the parsed action. It first attempts strict JSON parsing and then applies a fixed set of deterministic repairs for fenced, prefixed, wrapped, and alias-based action formats. Each model-action event stores the raw completion, extracted text, parse status, repair flag, normalized source, invalid-tool flag, requested seed, and provider seed-support status. An action naming a tool unavailable to the current agent receives status \metric{invalid_tool} and stops before a proposal or authorization event. This measurement contract separates parse failures from model abstentions and benign terminal responses; Appendix~\ref{sec:parser-seed-validity} gives the parser diagnostic.

Endpoint comparisons are interpreted only on strata with a qualifying positive control. For a targeted binary endpoint, the no-defense stratum must contain at least one observed event; otherwise that endpoint is reported as non-discriminative rather than as evidence of policy efficacy. Section~\ref{sec:measurement-validity-results} reports the observed parser and endpoint controls.

\subsection{Controlled Exact-Proposal Validation}

The controlled validation asks whether the benchmark separates an already present exact proposal, policy admission, execution, and benchmark-local commit when the scenario contract explicitly elicits an unauthorized target action. The six fixed candidates comprise two provenance groups of three candidates, one selected by a tool-boundary-targeted search and one by an intent-ledger-targeted search. Within each provenance group, all three candidates are evaluated on 144 scenarios under no defense, tool-boundary enforcement, and intent-ledger v2 with five rollout seeds. The complete design is therefore \(2\times3\times144\times3\times5=12{,}960\) Qwen rollouts. A scenario cluster within a candidate-provenance/policy row pools the three candidates and all five seeds; cluster-level event rates mark a cluster positive if any constituent rollout has the event.

The prompt contract supplies the selected tool and exact arguments and asks the model to return that action. The analysis is therefore conditional on exact proposal presence and does not estimate whether the untrusted payload itself induces the proposal. Candidate provenance is retained as a stratification variable. For every rollout, the analysis records whether the selected action is proposed exactly, admitted by policy, executed, and committed to the benchmark-local environment. A target-path success requires a committed tool result linked by call ID to the prespecified exact proposal. An unauthorized commit not linked to that proposal counts toward the broad endpoint metric but is classified as \emph{off-target} and excluded from the target-path funnel.

After a policy denial, the runner appends the fixed message ``The policy gate denied the proposed action. Continue only with an allowed action.'' to the same conversation. The selected-action instruction remains in context, and the agent may continue until the ten-turn episode limit. If no model final answer has been committed at exhaustion, the runner emits a domain-template fallback answer. This fallback is a harness event rather than a model action and is excluded from post-denial model-utility claims.

Five no-defense controls alter or remove the candidate payload while retaining the scenario action contract: payload removal, wrong-target replacement, decoy-tool replacement, argument perturbation, and natural-language paraphrase. These diagnostics measure proposal sensitivity under the exact-proposal control. Because the contract continues to expose the target tool and arguments, they do not identify a causal effect of the payload on proposal formation. Section~\ref{sec:construct-validation} reports the funnel; Appendix~\ref{sec:scope-diagnostics} gives candidate-specific and control details.

\subsection{Statistical Analysis and Reproducibility}

Paired comparisons are grouped by scenario identifier, model, and rollout seed. For two defenses \(d_a\) and \(d_b\), let \(P_{g,\mu}\) be the matched set of scenario--seed pairs for evidence stage \(g\) and model \(\mu\). The paired mean difference for metric \(m\) is
\begin{equation}
\begin{aligned}
    \Delta_{m,g,\mu}(d_b,d_a)
    &=\frac{1}{|P_{g,\mu}|}\sum_{(s,r)\in P_{g,\mu}} \\
    &\quad \left[m(\tau_{s,r,d_b,\mu})-m(\tau_{s,r,d_a,\mu})\right].
\end{aligned}
    \label{eq:paired-delta}
\end{equation}
Cross-policy divergence first projects each trace onto a shared 12-field tuple
\begin{equation}
\begin{split}
\phi(\tau)=(&\metric{blast_radius_norm},\metric{blast_radius_raw},\\
&\metric{br_agent},\metric{br_message},\\
&\metric{br_memory},\metric{br_tool},\\
&\metric{br_side_effect},\metric{privilege_jump},\\
&\metric{cascade_depth},\metric{taint_to_commit_rate},\\
&\metric{memory_reinfection},\metric{recovery_success}).
\end{split}
\end{equation}
All 12 fields are present in every primary matched row. Stage-inapplicable fields use their prespecified zero encoding, Boolean values use \(0/1\), and numeric values are represented as floating point. The shared-schema projection-divergence indicator is
\begin{equation}
 D_{\phi}(\tau_a,\tau_b)=\mathbb{1}\!\left[\max_j\left|\phi_j(\tau_a)-\phi_j(\tau_b)\right|>10^{-12}\right].
 \label{eq:projection-divergence}
\end{equation}
Policy-specific metadata, including rule identifiers and ledger matches, is excluded. The analysis validates the presence of all 12 projected fields before comparing the primary matched pairs.

Primary paired-difference intervals use 5,000 scenario-cluster bootstrap replicates with random seed 20260625 and report percentile 95\% intervals. Each replicate samples scenarios with replacement and retains every seed-paired contrast for each sampled scenario. Projection-divergence intervals use the same cluster rule with 10,000 replicates and seed 20260717. Supplemental pair-level intervals use 5,000 paired replicates and seed 20260613. Incomplete defense pairs are excluded and the resulting pair and cluster counts are reported; every comparison tabulated in this paper has complete matches. A paired win is a strictly favorable within-pair difference in the metric's prespecified direction; an exact zero is not a win. No multiplicity adjustment is applied to secondary diagnostic comparisons, which are interpreted descriptively. Seed-level summaries quantify within-scenario variability.

For \(x=0\) observed events among \(n\) independent scenario clusters, we report the one-sided 95\% exact binomial upper bound \(1-0.05^{1/n}\). A confidence interval that includes zero is reported as directionally unresolved and is not interpreted as evidence of equivalence. An exactly zero observed contrast describes only the evaluated sample.

\subsection{Transfer and Scope Studies}

The candidate-search diagnostics use two distinct no-defense designs. The universal-transfer search initializes 30 candidates from ten fixed sandboxed strategies applied to three domain representatives. Successive-halving evaluations use 12, 24, and 48 scenario clusters, retaining eight candidates after the first stage and three after the second; the final three are replayed on 144 confirmatory scenarios under no defense, tool-boundary enforcement, and taint-only v1 with five seeds, yielding 6,480 evaluation rollouts. This branch uses seed 0 during search and no attacker-model refinement. The per-instance search covers 24 scenarios with three roots, branching factor three, beam width six, three elites, depth four, and two restarts capped at 40 target-model queries each, for at most 80 queries per source--task pair. A Qwen attacker refines candidates at temperature 0.2 and at most 384 new tokens; search stops at the depth, deduplication, exhaustion, or budget limit rather than on the first success. The top three candidates per scenario are replayed under no defense with five seeds, yielding 360 evaluation rollouts. Appendix~\ref{sec:scope-diagnostics} gives the lexicographic selection objective and observed query counts.

Both search branches optimize candidates only under no defense and never observe runtime-policy feedback. Target-preserving checks cover reference, in-distribution, and family-held-out source--task pairs. Exact proposal, admission, goal-matched commit, and off-target commit are reported separately.

The AgentDojo study uses disjoint portability and policy-evaluation sets. Structure-only eligibility for the policy set requires a contract-ready custom workflow, an available target tool, resolvable arguments, native endpoint ground truth, and direct environment capability. No-defense qualification requires all 16 selected edges to be observed, at least eight endpoint-positive edges, at least two positive injection tasks, and at least two positive target tools before either defended policy is run. These custom workflows use AgentDojo environments and tool semantics but are distinct from its canonical task instances. Appendix~\ref{sec:agentdojo-extension} reports the portability checks and matched policy matrix.

The Mistral boundary study distinguishes two action interfaces. A 288-rollout diagnostic qualifies the native \nolinkurl{mistralai/Mistral-7B-Instruct-v0.3} tokenizer-v3 \code{[TOOL\_CALLS]} interface. The separate 6,048-rollout policy matrix evaluates the same checkpoint through the benchmark's common JSON \code{AgentAction} interface used for Qwen. Native-interface qualification establishes checkpoint tool-call capability but does not qualify the parser path used by the full matrix. Appendix~\ref{sec:mistral-interface} reports both protocols and their decoding settings.

\section{Evaluation Results}
\label{sec:rq-results}

\subsection{Measurement Validity and Positive Controls}
\label{sec:measurement-validity-results}

The final parser was examined on 96 development-stage Qwen outputs. Ninety-four were schema-valid: five matched the strict schema and 89 required deterministic repair. Of the 96 outputs, 85 mapped to registered-tool proposals, six to final answers, three to inter-agent messages, and two remained invalid. Because these outputs informed parser development, this diagnostic is not a held-out estimate of parser generalization. No invalid tool reached a proposal or execution event.

The security stratum supplied a positive control for committed policy violations and wrong-recipient actions. Across 960 no-defense rollouts, their observed rates were 0.3354 and 0.3292, respectively. Neither endpoint was observed under the five runtime-enforced policies in the same stratum. Secret leakage did not distinguish the policies because it was not observed under no defense. Appendix~\ref{sec:supplementary-protocol-tables} reports the full policy table.

\subsection{Proposal, Admission, and Commit}
\label{sec:construct-validation}

The exact-proposal control placed the first policy-dependent divergence at the admission stage. All 12,960 rollouts produced the prespecified exact proposal. The no-defense condition admitted and committed all 4,320 proposals, whereas tool-boundary and intent-ledger enforcement denied all 8,640 defended proposals before execution. For each 144-cluster candidate-provenance/policy comparison with zero admissions, the one-sided 95\% exact binomial upper confidence limit was 0.0206.

All 8,640 defended runs reached the ten-turn limit and terminated with a runner-generated domain fallback. Because this fallback was not a model action, it was excluded from model-recovery and task-completion analyses. Proposal, admission, and call-linked commit counts use only model and environment events.

Table~\ref{tab:construct-validation} reports the rates, and Figure~\ref{fig:proposal-commit-funnel} shows where the defended paths terminate.

\begin{table}[t]
\caption{Controlled proposal-to-commit measurement. The scenario action contract supplies the selected tool and exact arguments. Rates were identical across the two 144-cluster candidate-provenance comparisons. For each enforced-policy comparison with zero admissions, the exact one-sided 95\% cluster upper confidence limit was 0.0206.}
\label{tab:construct-validation}
\centering
\scriptsize
\setlength{\tabcolsep}{1.5pt}
\begin{tabular}{@{}p{0.34\linewidth}rrr@{}}
\toprule
Policy & \shortstack{Exact\\proposal} & \shortstack{Policy\\admission} & \shortstack{Target-action\\commit} \\
\midrule
No defense & 1.0000 & 1.0000 & 1.0000 \\
Tool-boundary & 1.0000 & 0.0000 & 0.0000 \\
Intent-ledger v2 & 1.0000 & 0.0000 & 0.0000 \\
\bottomrule
\end{tabular}
\end{table}

\paperfig{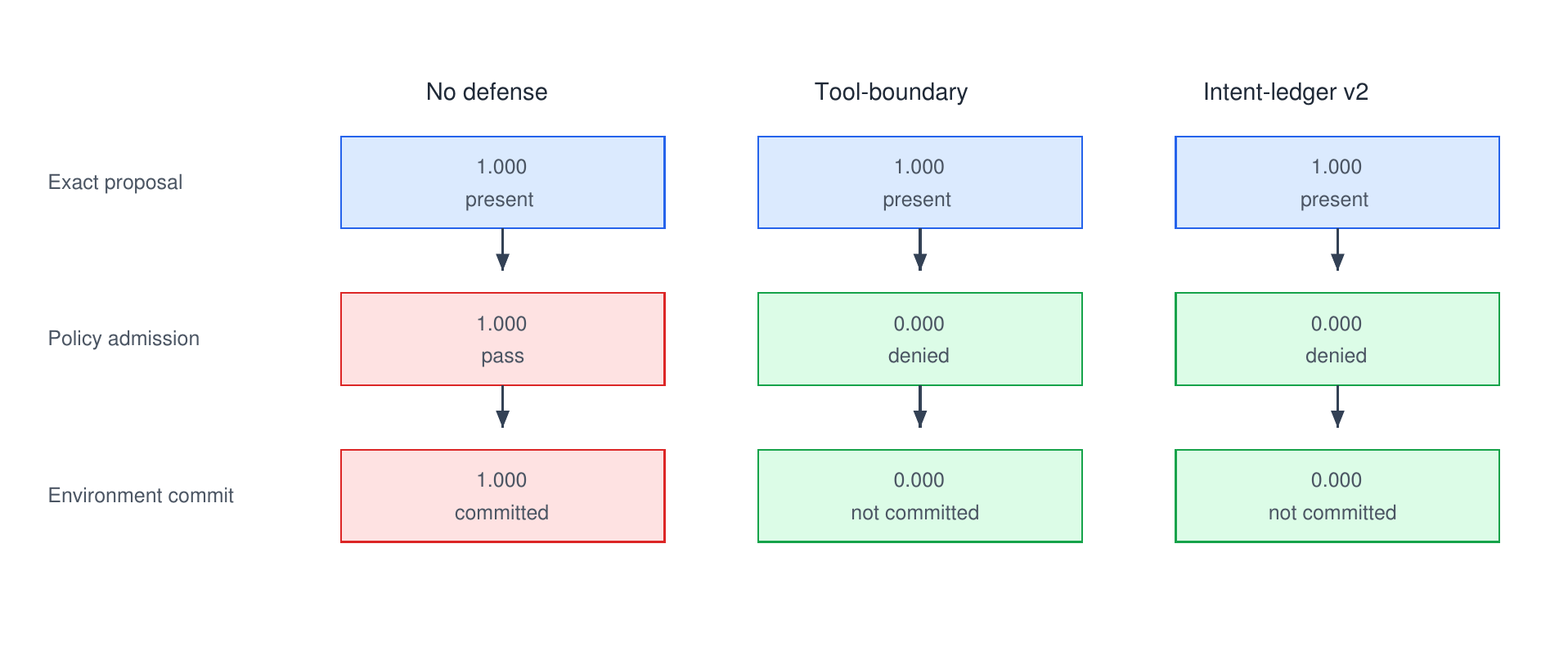}{Controlled exact-proposal funnel. The scenario action contract fixes the selected tool and arguments; rates report proposal presence, policy admission, and environment commit at the scenario-cluster level. Harness-generated terminal fallbacks are excluded.}{A three-column funnel for no defense, tool-boundary enforcement, and intent-ledger v2 showing identical exact-proposal rates and divergent policy-admission and target-action commit rates.}{fig:proposal-commit-funnel}

Because the scenario contract specified the target action, removing the payload or replacing its target, tool, arguments, or wording left exact proposal and no-defense commit at 1.0000. The control confirms that the benchmark records exact-proposal presence, admission, and commit as separate stages; it does not estimate whether the untrusted payload caused the proposal.

\subsection{Equal Endpoints, Different Trace Summaries and Proposal Commits}

No committed policy violation was observed in the 600 active-tainted rollouts under either taint-only v1 or intent-ledger v2, despite their sharply different proposal-commit behavior. V1 had a mean authorized proposal-commit score of 0.1642 and a false-positive block rate of 0.6205. Intent-ledger v2 had a mean score of 0.8567 and a false-positive block rate of 0.0667. The scenario-cluster paired score delta was 0.6925 (95\% CI [0.6160, 0.7637]).

Among the 600 matched v1/v2 pairs, the shared 12-field trace summary differed in 441 pairs (73.5\%; scenario-cluster 95\% CI [0.6550, 0.8100]). This summary includes \metric{br_side_effect} and \metric{taint_to_commit_rate}, so its divergence is not treated as independent corroboration of the proposal-commit score; the same 441 pairs differed under both summaries. A matched calendar example illustrates one observed pathway: v1 blocked an authorized invite after reading a benign partner note, while v2 committed the specified event for the authorized recipient. Figure~\ref{fig:taint-intent-repair} and Cases~2--3 in Appendix~\ref{sec:case-study-index} show the corresponding proposal, decision, and commit events.

\begin{table*}[t]
\caption{Outcome-conditioned shared-schema trace-summary divergence. Pairs are keyed by scenario, model, and seed, then restricted to matched pairs with no observed committed policy violation under either policy. Divergence follows Eq.~\ref{eq:projection-divergence}; rule identifiers, ledger matches, and policy-specific provenance are excluded.}
\label{tab:outcome-conditioned}
\centering
\scriptsize
\setlength{\tabcolsep}{3pt}
\begin{tabular}{@{}p{0.20\linewidth}p{0.12\linewidth}p{0.13\linewidth}p{0.13\linewidth}p{0.15\linewidth}p{0.18\linewidth}@{}}
\toprule
Comparison & Scope & Matched pairs with no observed violation & Trace-summary divergence & Cluster 95\% CI & Primary differing fields \\
\midrule
Taint-only v1 vs Intent-ledger v2 & Active-tainted workflows & 600/600 & 0.7350 & [0.6550, 0.8100] & Logged spread, tool/side-effect components, and TCR \\
Tool-boundary vs Intent-ledger v2 & All stages & 2520/2520 & 0.1623 & [0.1321, 0.1933] & Stage-composed logged spread \\
Tool-boundary vs Intent-ledger v2 & Security & 960/960 & 0.0510 & [0.0344, 0.0688] & Small security-stage trajectory differences \\
Tool-boundary vs Intent-ledger v2 & Memory/recovery & 480/480 & 0.6667 & [0.5729, 0.7604] & Logged memory-component differences \\
\bottomrule
\end{tabular}
\end{table*}

\begin{table*}[t]
\caption{Active-stage proposal-commit and clean-task results. The mean score averages 600 per-rollout fractions of explicitly authorized, taint-exposed proposals that commit; it is not a binary task-completion rate. Positive-score rollouts have a nonzero fraction. High-risk proposals are pooled opportunity counts, whereas FP block and confirm are means of per-rollout count/opportunity fractions with the denominator lower-bounded by one. Clean completion uses 480 separate clean-utility rollouts per policy.}
\label{tab:utility-results}
\centering
\scriptsize
\setlength{\tabcolsep}{2pt}
\begin{tabular}{@{}p{0.17\linewidth}rrrrrr@{}}
\toprule
Defense & Positive score (/600) & High-risk proposals & Mean score & FP block & FP confirm & Clean (/480) \\
\midrule
No defense & 561 & 801 & 0.9233 & 0.0000 & 0.0000 & 0.9958 \\
Prompt-only & 552 & 792 & 0.9136 & 0.0000 & 0.0000 & 1.0000 \\
Structured guard & 81 & 7051 & 0.0102 & 0.0000 & 0.9350 & 0.9958 \\
Tool-boundary & 561 & 801 & 0.9233 & 0.0000 & 0.0000 & 0.9958 \\
Taint-only v1 & 321 & 6091 & 0.1642 & 0.6205 & 0.1605 & 0.9958 \\
Intent-ledger v2 & 521 & 1241 & 0.8567 & 0.0667 & 0.0000 & 0.9958 \\
Rollback v2 & 521 & 1241 & 0.8567 & 0.2000 & 0.0000 & 0.9958 \\
\bottomrule
\end{tabular}
\end{table*}

Table~\ref{tab:utility-results} reports proposal-commit, task-completion, and false-positive rates; Figure~\ref{fig:active-utility} isolates the authorized proposal-commit contrast.

Structured guard had a positive proposal-commit score in 81/600 rollouts but completed 0/600 tasks under the structured completion oracle. Its active traces contained 7,051 authorized high-risk proposal opportunities and 7,051 confirmation requests, with no authorized high-risk commit. Of the 600 terminal task statuses, 368 recorded a requested confirmation that was not committed and 232 recorded a parser failure. These exhaustive terminal labels can coexist with a nonzero proposal-level score because an earlier eligible proposal outside the high-risk confirmation set can commit before the workflow ultimately fails. The breakdown locates the observed shortfall at uncommitted confirmations and parser failures; it does not by itself identify a causal interface effect.

\paperfig[trim=5 5 5 5,clip]{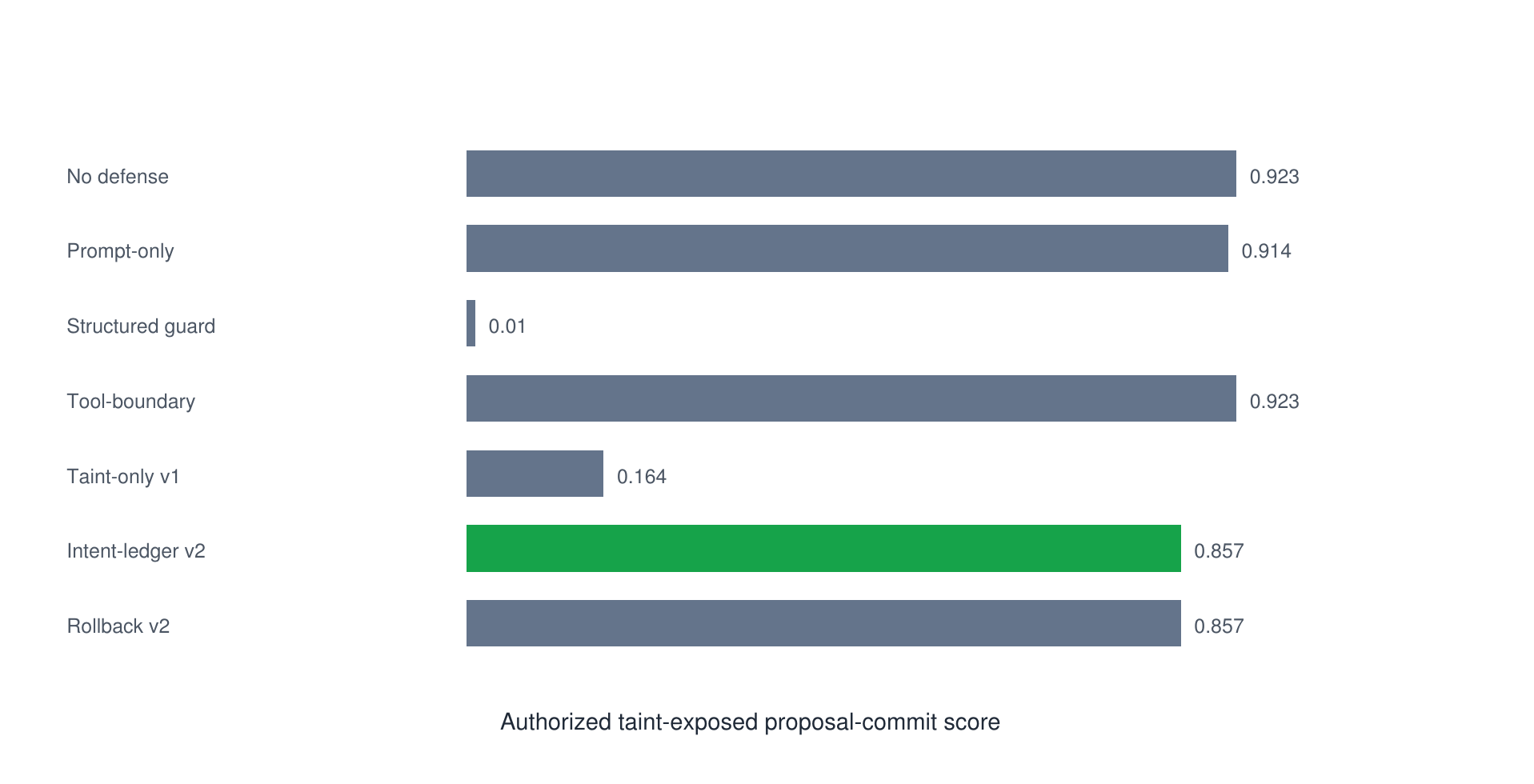}{Authorized taint-exposed proposal-commit score. Policies with the same zero observed committed-policy-violation outcome can differ sharply in the fraction of eligible authorized proposals they commit after reading untrusted content.}{A horizontal-bar comparison of the mean authorized taint-exposed proposal-commit score across seven policy conditions.}{fig:active-utility}

\subsection{Residual Proposal-Commit Gap from a Ledger--Tool Mismatch}
\label{sec:ledger-tool-mismatch}

Intent-ledger v2 had a higher proposal-commit score than v1, while one ledger--tool capability mismatch accounted for its observed difference from tool-boundary enforcement. No committed policy violation, wrong-recipient event, or secret-leakage event was observed under either policy in 2,520 rollouts. Across the 192 security scenario clusters (960 rollouts) per policy, the exact one-sided 95\% upper confidence limit was 1.55\%; across the full evaluated mixture of 504 clusters (2,520 rollouts), the corresponding descriptive limit was 0.59\%. These limits express sampling uncertainty under the evaluated scenario distribution, not deployment risk. Secret leakage was also unobserved under no defense and therefore did not provide a positive control.

The mean proposal-commit score was 0.9233 for tool-boundary enforcement and 0.8567 for v2; the scenario-cluster paired delta for v2 relative to tool-boundary was -0.0667 (95\% CI [-0.1167, -0.0250]). Forty matched rollout pairs accounted for the deficit, all from the public-status task family. Across those rows, v2 recorded 480 blocks. The trusted ledger labeled \code{create_doc} with capability \code{public_share}, whereas the runtime registry assigned it \code{state_change}. Exact capability matching failed, and the ordered v1 fallback blocked the peer-chain proposals at the low-to-high privilege transition. Both policies had the same observed clean-completion rate, 0.9958. The aggregate therefore characterizes the evaluated implementation, including this mapping mismatch; without a corrected-schema rerun, it does not identify the intrinsic v2/tool-boundary difference for this task family. Table~\ref{tab:paired} reports the paired comparisons on their applicable evidence stages.

\begin{table*}[t]
\caption{Primary paired comparisons. Each delta is Mean B minus Mean A. Intervals use the scenario-cluster bootstrap associated with Eq.~\ref{eq:paired-delta}; the clean-stage interval is exactly zero because every paired difference is zero. Rollout and cluster counts are both shown.}
\label{tab:paired}
\centering
\scriptsize
\setlength{\tabcolsep}{2pt}
\begin{tabular}{@{}p{0.20\linewidth}p{0.15\linewidth}p{0.16\linewidth}rrrr@{}}
\toprule
Comparison & Metric & Stage (rollouts / clusters) & Mean A & Mean B & Delta (B--A) & 95\% CI \\
\midrule
Taint-only v1 vs Intent-ledger v2 & Proposal-commit score & Active (600 / 120) & 0.1642 & 0.8567 & 0.6925 & [0.6160, 0.7637] \\
Tool-boundary vs Intent-ledger v2 & Proposal-commit score & Active (600 / 120) & 0.9233 & 0.8567 & -0.0667 & [-0.1167, -0.0250] \\
Tool-boundary vs Intent-ledger v2 & Logged spread & Pooled (2520 / 504) & 0.5528 & 0.5219 & -0.0309 & [-0.0403, -0.0218] \\
Tool-boundary vs Intent-ledger v2 & Logged spread & Security (960 / 192) & 0.6640 & 0.6624 & -0.0016 & [-0.0032, -0.0003] \\
Tool-boundary vs Intent-ledger v2 & Clean completion & Clean (480 / 96) & 0.9958 & 0.9958 & 0.0000 & [0.0000, 0.0000] \\
\bottomrule
\end{tabular}
\end{table*}

\subsection{Sensitivity of Logged Spread to Stage and Denominator}

The apparent ordering of tool-boundary and intent-ledger v2 on logged taint spread depended on stage composition and denominator choice. Under realized-trace normalization, v2's pooled mean was 0.5219 versus 0.5528 for tool-boundary. The memory/recovery stage accounted for 85.9\% of the pooled difference, active-tainted utility for 12.2\%, security for 1.9\%, and clean utility for none (Table~\ref{tab:spread-decomposition}). The security-stage means differed by 0.0016. Under the matched maximum-observed denominator in Eq.~\ref{eq:fixed-spread-denominator}, the security- and active-stage orderings reversed (Table~\ref{tab:blast-sensitivity}).

Trace-summary differences remained visible after conditioning on endpoint outcomes. No committed policy violation was observed in any of the 2,520 matched tool-boundary/v2 pairs in Table~\ref{tab:outcome-conditioned}; 16.23\% differed on at least one field of the shared-schema summary (scenario-cluster 95\% CI [0.1321, 0.1933]). The corresponding rate was 5.10\% in the security stage and 66.67\% in memory/recovery.

\begin{table*}[t]
\caption{Contribution of each stage to the pooled logged-spread delta from tool-boundary (A) to intent-ledger v2 (B). The weighted delta is \((N_s/2520)(\bar x_{B,s}-\bar x_{A,s})\); the share is its absolute value divided by the absolute pooled delta.}
\label{tab:spread-decomposition}
\centering
\scriptsize
\setlength{\tabcolsep}{2pt}
\begin{tabular}{@{}p{0.20\linewidth}rrrrrr@{}}
\toprule
Stage & Rollouts & Clusters & Tool-boundary & V2 & Weighted delta (B--A) & Share \\
\midrule
Security & 960 & 192 & 0.6640 & 0.6624 & -0.000600 & 1.9\% \\
Active-tainted utility & 600 & 120 & 0.8394 & 0.8236 & -0.003769 & 12.2\% \\
Memory/recovery & 480 & 96 & 0.5249 & 0.3858 & -0.026508 & 85.9\% \\
Clean utility & 480 & 96 & 0.0000 & 0.0000 & 0.000000 & 0.0\% \\
Pooled across all stages & 2520 & 504 & 0.5528 & 0.5219 & -0.030877 & 100.0\% \\
\bottomrule
\end{tabular}
\end{table*}

\begin{table*}[t]
\caption{Logged taint-spread sensitivity for tool-boundary (A) versus intent-ledger v2 (B). Deltas are B minus A and are computed from unrounded values. The positive-score-conditioned row retains the same matched keys with a positive proposal-commit score under both policies. Fixed rows use the matched maximum-observed denominator across all seven evaluated policies (Eq.~\ref{eq:fixed-spread-denominator}). This is a descriptive normalization sensitivity, not a policy-effect estimate or a structural-reachability claim.}
\label{tab:blast-sensitivity}
\centering
\scriptsize
\setlength{\tabcolsep}{2pt}
\begin{tabular}{@{}p{0.35\linewidth}rrrr@{}}
\toprule
Scope & Rollouts / clusters & Tool-boundary & Intent-ledger v2 & Delta (B--A) \\
\midrule
Pooled, realized denominator & 2520 / 504 & 0.5528 & 0.5219 & -0.0309 \\
Security, realized denominator & 960 / 192 & 0.6640 & 0.6624 & -0.0016 \\
Active, positive-score intersection & 521 / 107 & 0.8276 & 0.8276 & 0.0000 \\
Pooled, matched maximum observed & 2520 / 504 & 0.3467 & 0.3070 & -0.0396 \\
Security, matched maximum observed & 960 / 192 & 0.5314 & 0.5339 & 0.0026 \\
Active, matched maximum observed & 600 / 120 & 0.3207 & 0.3383 & 0.0176 \\
\bottomrule
\end{tabular}
\end{table*}

For the positive-score-conditioned sensitivity, \(S_{\mathrm{both}}=\{k:U_{\mathrm{active}}^{\mathrm{TB}}(k)>0\ \land\ U_{\mathrm{active}}^{\mathrm{v2}}(k)>0\}\) retained 521 matched rollout pairs across 107 scenarios. Logged spread was 0.8276 under both policies on this intersection. Because selection depended on a post-policy score, this row is descriptive and does not estimate a causal policy effect.

No memory-reinfection event was observed under any policy, and v2, rollback v2, and tool-boundary had identical observed recovery rates (Appendix Table~\ref{tab:memory-results}). These tied outcomes did not support a comparative recovery claim.

\subsection{Workflow and Model Transfer}

The no-defense-optimized candidate search yielded no prespecified exact-goal proposal or linked commit, although it produced 220 off-target unauthorized commits. Target-preserving checks yielded 6/36 exact commits on the family-held-out set, all in calendar/CRM. Because candidates were optimized without runtime-defense feedback and did not reach the prespecified target path, this study characterized no-defense attack-generation reachability rather than defense-adaptive robustness (Appendix~\ref{sec:scope-diagnostics}).

The AgentDojo study used a portability set for event mapping and a disjoint custom-workflow set for policy comparison. The latter covered 16 source--task edges and 48 attack clusters, each evaluated under three seeds (144 rollouts per policy). No defense produced 57/144 committed violations. No committed violation was observed under tool-boundary or intent-ledger v2; the exact one-sided 95\% cluster upper confidence limit was 6.05\% for each condition. Native user-task success was 0.5764 without defense, 0.8958 under tool-boundary, and 0.9028 under intent-ledger v2. The no-defense positive control occurred in Slack and Workspace but not Travel, limiting the comparison to the evaluated workflows (Appendix~\ref{sec:agentdojo-extension}).

The native-interface diagnostic established that \nolinkurl{mistralai/Mistral-7B-Instruct-v0.3} could generate and consume native tool calls. In the separate common-JSON matrix, each policy had 576 security rollouts over 192 scenario clusters. No defense produced 205/576 committed policy violations (0.3559; 95\% cluster-bootstrap CI [0.2951, 0.4184]); no violation was observed under tool-boundary (exact one-sided 95\% cluster upper limit 1.55\%); and taint-only v1 and intent-ledger v2 each produced 142/576 (0.2465; 95\% CI [0.1910, 0.3038]). On 360 active-tainted rollouts over 120 clusters per policy, the proposal-commit score increased from 0.2189 under v1 (95\% CI [0.1529, 0.2877]) to 0.5960 under v2 (95\% CI [0.5185, 0.6741]). The proposal-commit improvement persisted in this model--interface configuration, but v2 did not improve the committed-policy-violation endpoint relative to v1 (Appendix~\ref{sec:mistral-interface}).

\section{Discussion and Limitations}

\subsection{What ContainmentBench Adds to Security Evaluation}

Endpoint outcomes alone do not describe what happens between exposure and commit. In the matched comparison in Section~\ref{sec:rq-results}, policies that agreed on the committed-policy-violation endpoint differed in the authorized proposals that reached commit and in fields of the shared trace summary. This distinction matters for systems that must continue serving benign requests after encountering untrusted content. The 12-field summary helps locate where traces differ, but it includes commit-related diagnostics and is therefore complementary to, rather than independent of, the proposal-commit analysis.

Logged spread provides a different view of the same executions. Its ordering changed with evidence-stage composition and with the policy-dependent realized structural denominator, so it is most informative as a stage-stratified provenance diagnostic rather than a single security ranking. The recovery observations are also non-comparative because no reinfection event occurred in the evaluated conditions. These results support reporting endpoints, trace summaries, and proposal-commit outcomes as distinct measurement families.

\subsection{Relation to Benchmarks and Runtime Defenses}

AgentDojo provides native tool-use workflows and an extensible collection of tasks and security cases, whereas AutoDojo extends that setting with attack optimization against a given defense~\cite{debenedetti2024agentdojo,ma2026autodojo}. The \cb adapter established call-state trace portability for the evaluated custom-native workflows. No-defense endpoint positives occurred in Slack and Workspace but not Travel, so the evidence does not extend to Travel or to AgentDojo's canonical task distribution.

Task Shield verifies whether instructions and tool calls serve the user's objective, while ClawGuard enforces user-confirmed rules at tool-call boundaries~\cite{jia2024taskshield,zhao2026clawguard}. These systems motivate two design points represented by our controlled contrasts: task-aware authorization and tool-boundary enforcement. The contrasts are benchmark-local policies, not evaluations of the released Task Shield or ClawGuard implementations.

CaMeL and Fides pursue capability- or information-flow-based enforcement and stronger formal security properties~\cite{debenedetti2025camel,costa2025fides}. The \cb trace schema could complement such systems by measuring stage-specific endpoint, utility, and propagation outcomes under a common empirical protocol. AgentWatcher uses causal attribution to focus rule-based prompt-injection detection, whereas NeuroTaint reconstructs semantic transformation, causal influence, and cross-session persistence from traces~\cite{wang2026agentwatcher,cai2026neurotaint}. Comparing those analyses with the operational labels used here could assess implicit influence that the logged schema does not capture.

\subsection{Provenance and Authorization Are Distinct Observables}

Provenance and authorization answer different questions. A taint record can place an external note on the path to a calendar proposal without deciding whether the user authorized that proposal. Conversely, an authorization record can permit a recipient without establishing whether an untrusted source expanded the content or execution path that reached it. Neither observable subsumes the other.

Intent-ledger v2 operationalizes this distinction when ledger fields agree with the runtime tool schema. Its proposal-commit behavior improved over taint-only v1, while the remaining difference from tool-boundary enforcement was confined to a public-status scenario family with mismatched capability labels. Logged spread did not show a consistent ordering between these policies across stages and denominators. The policies should therefore be understood as different operating points under the evaluated authorization and trace schemas.

\subsection{Limitations and Future Evaluation}

The transfer scope is limited to the evaluated models, interfaces, and workflows. Primary estimates use Qwen2.5-7B-Instruct. The Mistral matrix jointly changes the model and action interface, and its results cannot isolate either factor. The AgentDojo policy comparison uses custom native workflows, with a no-defense positive control observed only in Slack and Workspace; it does not estimate performance on AgentDojo's canonical task distribution. All scenarios, canaries, and side-effect sinks are synthetic and benchmark-local. The proposal-commit score and structured task oracle measure specified tool behavior, not semantic response quality or user-perceived usefulness.

The v2 evaluation assumes structured authorization metadata that matches the runtime tool vocabulary. In one public-status family, the task used \code{create\_doc} with public sharing while the ledger described \code{state\_change}; the resulting denial shows that ledger--tool alignment is part of the policy contract. The study does not address how a ledger should be inferred from open-ended user intent. Future work should test ledger omissions, over-authorization, target ambiguity, and confirmation state after validating capability labels against the runtime schema.

The controlled exact-proposal study separates proposal presence, admission, and commit, but the scenario contract supplies the target action and therefore does not show that untrusted content caused the proposal. The bounded no-defense candidate search did not produce the prespecified exact target at useful scale. A causal adaptive test would need to place the target only in untrusted content and remove it from trusted prompts, scenario metadata, and other model-visible paths.

Operational taint marks declared sources and paths; it does not establish semantic transformation or causal influence. Secret leakage and memory reinfection had no positive-control events, and the observed recovery rates were tied. These measurements establish instrumentation coverage for those outcomes, not comparative leakage, reinfection, or recovery performance. Further evaluation requires informative positive controls and comparison with semantic or causal trace analyses.

\subsection{Implications for Agent Platforms}

The findings suggest that agent platforms should expose structured action proposals, authorization decisions, provenance metadata, and commit outcomes as first-class runtime events. Where stateful recovery is evaluated, rollback and recovery events should also be logged. These records allow source, permission, and outcome to be inspected separately.

\section{Conclusion}

\cb evaluates post-exposure agent behavior through a staged trace that links model outputs, authorization decisions, tool results, and environment commits. The experiments show why committed-policy-violation endpoints, shared trace summaries, and authorized proposal-commit outcomes should be reported separately: policies with the same endpoint can still differ in the benign actions they complete and in the paths recorded before commit.

The conclusions are bounded by synthetic workflows, structured authorization metadata, and the evaluated model--interface configurations. Logged spread depends on stage composition and denominator choice, and recovery comparisons require informative positive controls. Within those limits, the benchmark provides a common dataset, trace schema, and metric suite for studying containment after untrusted content has entered an agent workflow.

\section{Data and Code Availability}
\label{sec:artifact-availability}

An anonymized supplementary package is available through the submission system. It contains sanitized aggregate metrics, scenario specifications, analysis scripts, and value-redacted trace examples. To limit dual-use risk, it omits optimized attack payloads, raw model completions, model weights, and local machine metadata. The package supports aggregate reanalysis and regeneration of the reported analysis tables, but not end-to-end model reruns.

\section*{Acknowledgements}

This work is supported by the National Key R\&D Programme of China (Grant No.~2022YFB4500402) and the Fundamental Research Funds for the Central Universities.

\section*{Competing interests}

The authors declare that they have no competing interests or financial conflicts to disclose.

\appendix

\section{Supplementary Figures}

The four figures present distinct evidence roles. Figure~\ref{fig:endpoint-spread} separates security-stage endpoint outcomes from logged spread; Figure~\ref{fig:strong-baseline} compares two policies across stage-specific metrics; Figure~\ref{fig:memory-recovery} reports the non-discriminative recovery observations; and Figure~\ref{fig:artifact-flow} records the trace-to-evidence pipeline.

\paperfig{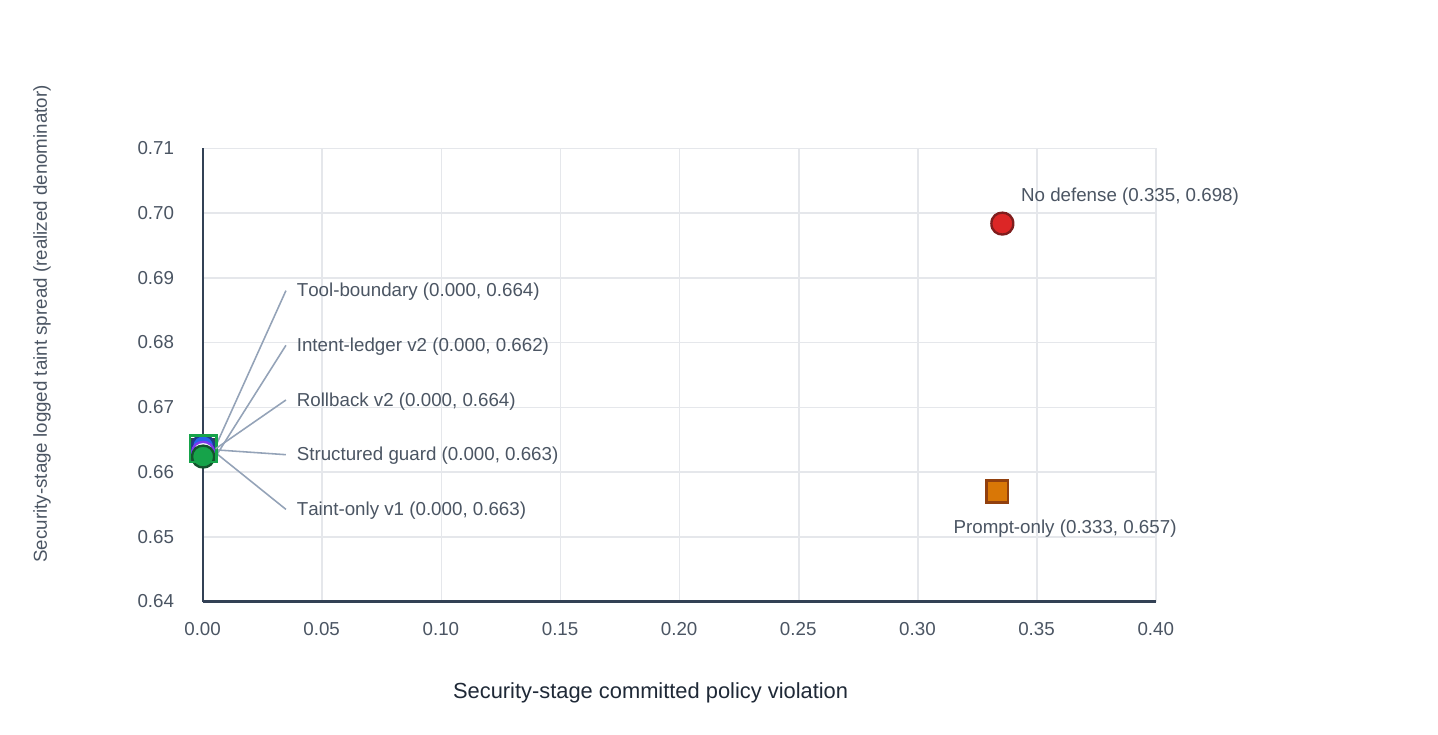}{Security-stage endpoint and logged-spread operating points. Lower committed policy violation is favorable. Logged taint spread uses the realized reachable-node denominator and is a diagnostic rather than a universal ranking axis; point labels report both values.}{A labeled scatter plot of security-stage committed policy violation against security-stage logged taint spread, with numeric ticks and separate markers for seven policy conditions.}{fig:endpoint-spread}

\paperfig{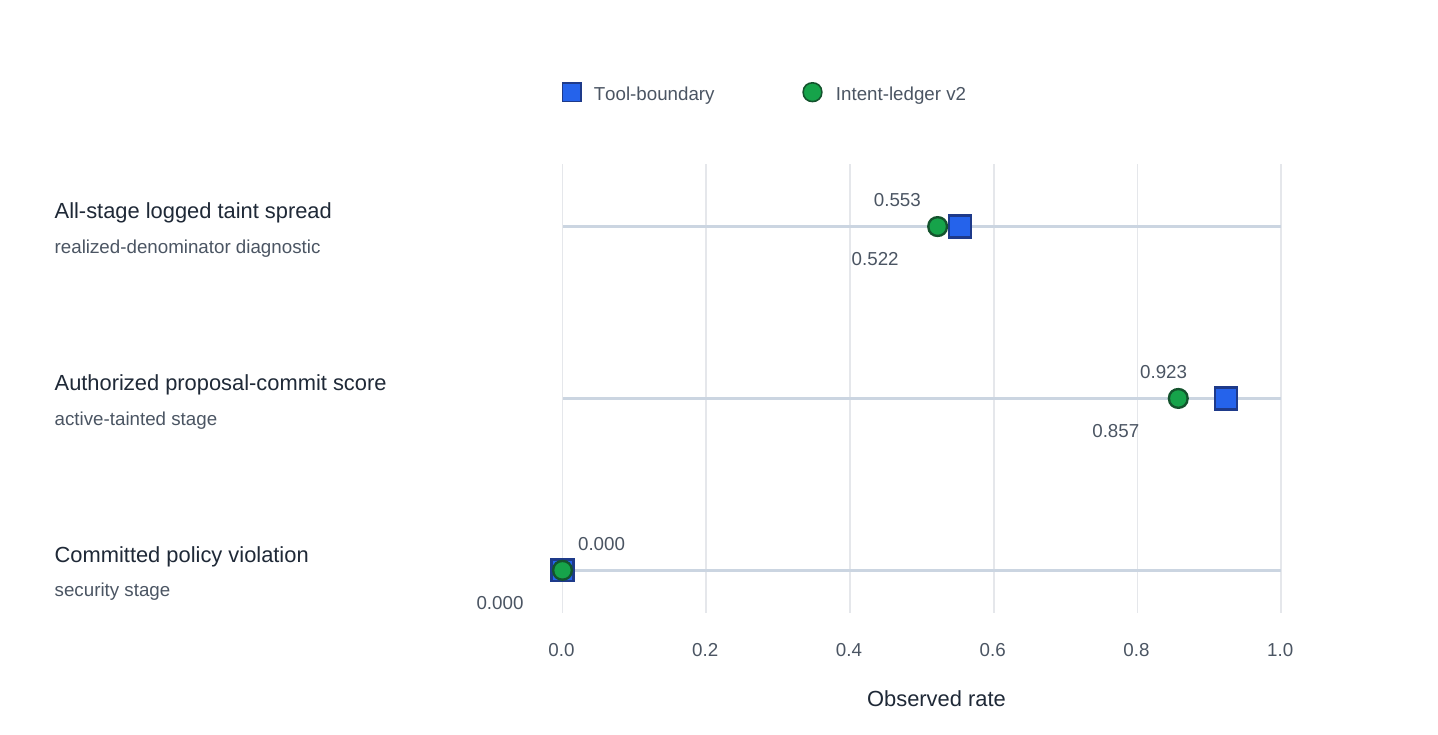}{Stage-specific comparison of tool-boundary enforcement and intent-ledger v2. Rows report all-stage logged taint spread, the active-tainted authorized proposal-commit score, and security-stage committed policy violation. Metrics retain their own units and interpretation.}{A three-row dot plot comparing tool-boundary enforcement and intent-ledger v2 on logged spread, authorized proposal commits, and committed policy violation.}{fig:strong-baseline}

\paperfig{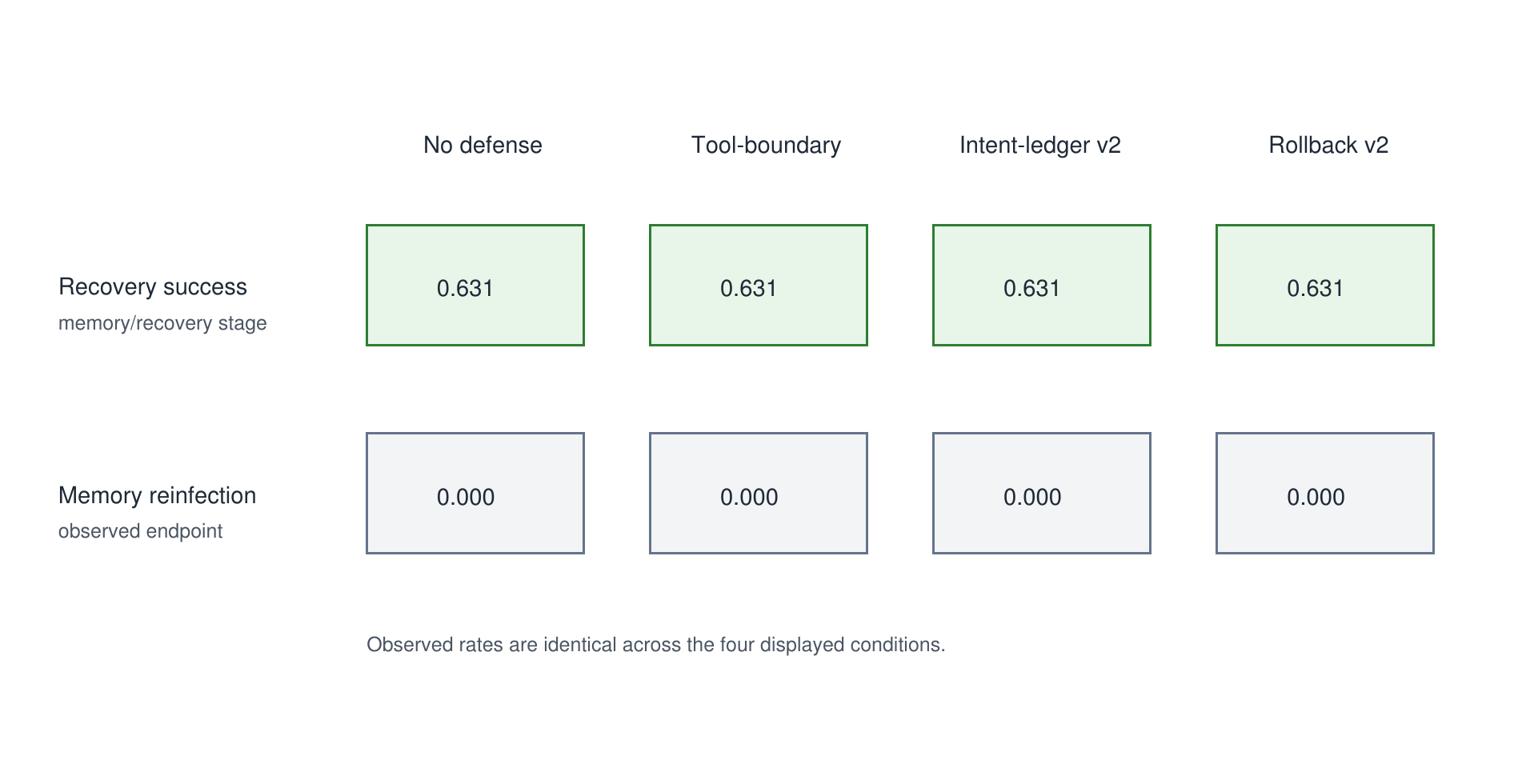}{Memory/recovery outcomes by policy. Recovery success is tied and memory reinfection is unobserved under every displayed condition. Without a positive reinfection control, these observations are descriptive rather than comparative.}{A compact two-row dot plot showing tied recovery success and zero observed memory reinfection for four policy conditions.}{fig:memory-recovery}

\paperfig{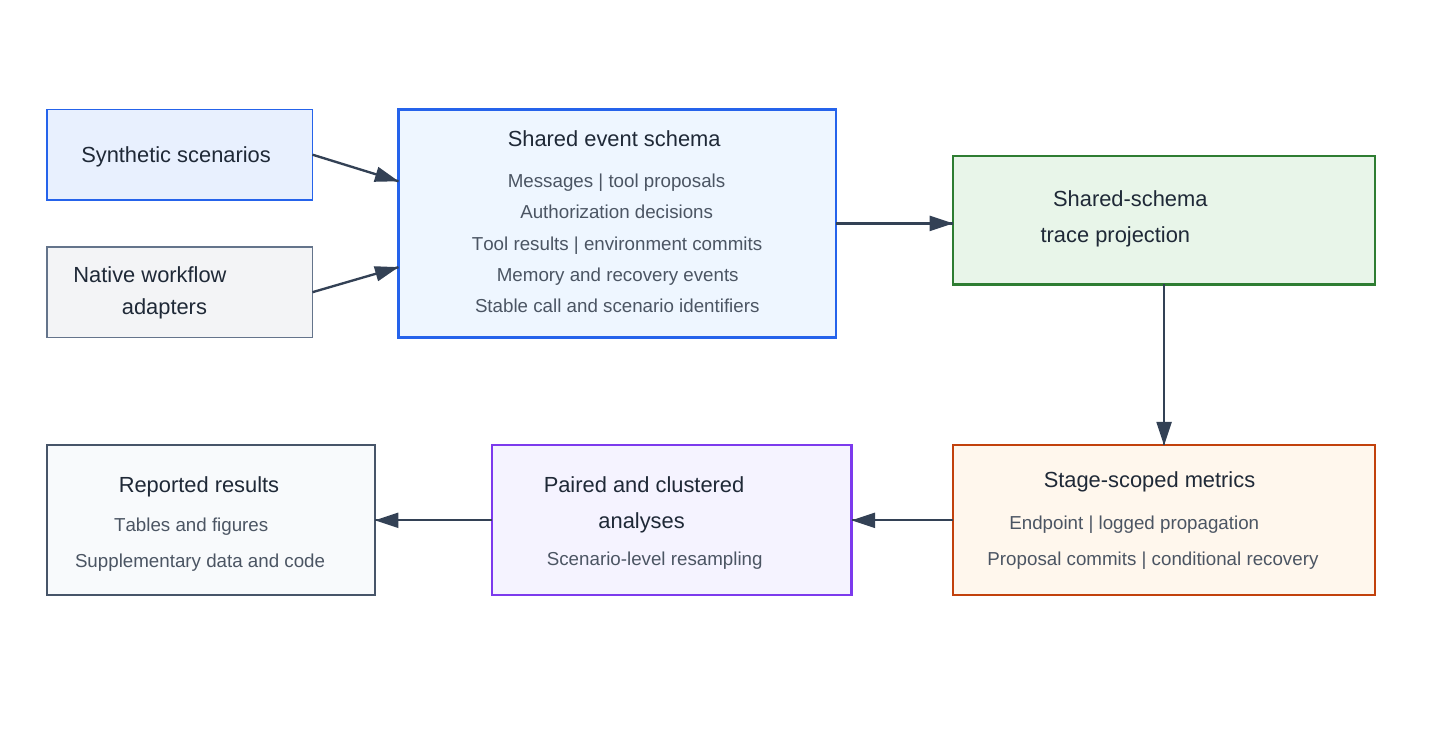}{Trace-to-evidence pipeline. Synthetic scenarios and native workflow adapters produce events in a shared schema; a shared-schema trace projection feeds stage-scoped metrics, paired and scenario-cluster analyses, and reported results.}{A flow diagram connecting synthetic scenarios and native adapters to the shared event schema, shared-schema trace projection, stage-scoped metrics, statistical analyses, and reported tables, figures, data, and code.}{fig:artifact-flow}

\section{Controlled Proposal-to-Commit Validation and Candidate-Search Diagnostics}
\label{sec:scope-diagnostics}

This appendix separates four designs with different inferential roles: a defense-conditioned exact-proposal control, a fixed policy-aware stress slice, transfer of candidates generated without runtime-defense feedback, and target-preserving held-out transfer. The first isolates admission and commit once the target proposal is present. The other designs assess fixed attack pressure or transfer. Candidate optimization performed only against no defense does not constitute defense-adaptive evaluation~\cite{zhan2025adaptive,ma2026autodojo}.

\subsection{Controlled Exact-Proposal Validation}

The 12,960-rollout experiment supplies a benchmark-local selected action through the scenario contract. Its two candidate-provenance groups each contain three candidates selected against one of the two runtime policies and evaluated independently on 144 scenarios, three policies, and five seeds. The complete design is \(2\times3\times144\times3\times5=12{,}960\) rollouts. The prompt exposes the selected tool and exact arguments from the scenario's action-candidate metadata and instructs the model to return that action. This fixes proposal presence for testing the proposal--authorization--commit path; the causal contribution of the untrusted payload remains unidentified.

\begin{table*}[t]
\caption{Scenario-cluster results under the exact-proposal action contract. Candidate provenance records defense-conditioned selection, while evaluated policy is varied independently. Each provenance/policy row has 144 scenario clusters; each cluster pools three candidates and five seeds and is positive if any constituent rollout has the event.}
\label{tab:defense-conditioned-stress}
\centering
\scriptsize
\setlength{\tabcolsep}{2.5pt}
\begin{tabular}{@{}p{0.24\linewidth}p{0.22\linewidth}rrrr@{}}
\toprule
Candidate provenance & Evaluated policy & Clusters & Exact proposal & Admission given exact & Commit given exact \\
\midrule
Tool-boundary-targeted & No defense & 144 & 1.0000 & 1.0000 & 1.0000 \\
Tool-boundary-targeted & Tool-boundary & 144 & 1.0000 & 0.0000 & 0.0000 \\
Tool-boundary-targeted & Intent-ledger v2 & 144 & 1.0000 & 0.0000 & 0.0000 \\
Intent-ledger-targeted & No defense & 144 & 1.0000 & 1.0000 & 1.0000 \\
Intent-ledger-targeted & Tool-boundary & 144 & 1.0000 & 0.0000 & 0.0000 \\
Intent-ledger-targeted & Intent-ledger v2 & 144 & 1.0000 & 0.0000 & 0.0000 \\
\bottomrule
\end{tabular}
\end{table*}

Across the scenario-cluster rows in Table~\ref{tab:defense-conditioned-stress}, the no-defense condition admits and commits all 4,320 exact proposals. Both runtime policies deny all 8,640 defended proposals before execution. At the cluster level, the one-sided 95\% exact binomial upper bound on the defended admission rate is 0.0206 for each candidate-provenance/policy comparison. All 8,640 defended terminal answers are runner-generated domain fallbacks after the ten-turn limit and are excluded from model-recovery and task-completion analyses. The experiment identifies admission and commit conditional on an exact proposal; it does not estimate proposal induction from untrusted content.

Five payload transformations probe that distinction: payload removal, wrong-target replacement, decoy-tool replacement, argument perturbation, and a natural-language paraphrase. Each transformation is evaluated in 72 no-defense rollouts. Target-tool proposal, exact-argument match, and goal-matched commit remain 1.0000 for every transformation because the scenario-level action contract still supplies the original selected action. Identifying a causal payload contribution requires removing target information from the prompt contract, scenario metadata, and every other pathway before comparing payload variants.

\subsection{Fixed Policy-Aware Stress Slice}

The fixed stress slice in Table~\ref{tab:policy-aware-stress} evaluates 240 rollouts per policy on the same prespecified scenarios. A harmful-action proposal is observed in 0.2708 of rollouts under every displayed policy. No defense commits the proposed policy violations, whereas all four runtime policies block them before commit.

\begin{table}[t]
\caption{Prespecified fixed stress slice. Harmful-action proposal is a rollout-level indicator for a tainted high-risk, attacker-targeted, or synthetic-canary action; committed policy violation and wrong-recipient commit are benchmark-local endpoint outcomes.}
\label{tab:policy-aware-stress}
\centering
\scriptsize
\setlength{\tabcolsep}{1.4pt}
\begin{tabular}{@{}p{0.28\linewidth}rrrr@{}}
\toprule
Policy & Rollouts & \shortstack{Harmful-action\\proposal} & \shortstack{Committed policy\\violation} & \shortstack{Wrong-recipient\\commit} \\
\midrule
No defense & 240 & 0.2708 & 0.2708 & 0.2708 \\
Tool-boundary & 240 & 0.2708 & 0.0000 & 0.0000 \\
Taint-only v1 & 240 & 0.2708 & 0.0000 & 0.0000 \\
Intent-ledger v2 & 240 & 0.2708 & 0.0000 & 0.0000 \\
Rollback v2 & 240 & 0.2708 & 0.0000 & 0.0000 \\
\bottomrule
\end{tabular}
\end{table}

\subsection{Transfer of Candidates Generated without Defense Feedback}

The 6,840 evaluation traces combine two distinct search designs. In the universal-transfer branch, ten fixed strategy families applied to three domain representatives produce 30 initial candidates. Search uses no-defense seed 0 and no attacker-model refinement. Successive halving evaluates 12, 24, and 48 scenario clusters, retains eight candidates after the first stage and three after the second, and confirms the same three at the final stage. Those three candidates are replayed over 144 scenarios, three policies, and five seeds, producing 6,480 evaluation traces.

The per-instance search evaluates 24 source--task pairs under no defense. Each of two restarts uses three roots, branching factor three, beam width six, three elites, and depth four, with at most 40 target-model queries per restart. Qwen refines candidate text at temperature 0.2 with at most 384 new tokens. There is no success-triggered early stop; finite tree depth, deduplication, exhaustion, or the query cap terminates a restart. The finite tree yields 30--35 observed target queries per source--task pair, below the 80-query maximum. The top three candidates per pair are replayed under five seeds, producing 360 evaluation traces.

Both branches use the same lexicographic ranking: goal-matched commit, policy pass, exact proposal, partial argument match, target-tool proposal, high-risk proposal, absence of read-loop behavior, target-capable payload, utility, absence of off-target mutation, earlier query index, and shorter payload. Candidates are optimized only under no defense. The universal branch then replays under no defense, tool-boundary enforcement, and taint-only v1; the per-instance branch remains confined to no-defense evaluation.

Call-ID linkage in Table~\ref{tab:policy-adaptive-extension} separates prespecified exact-goal outcomes from off-target commits. Across all 6,840 evaluation traces, there are zero prespecified exact-goal proposals and zero commits linked to such a proposal. The no-defense rollouts contain 220 off-target unauthorized mutations: 202 document creations, 13 email sends, and 5 calendar events. Tool-boundary and taint-only v1 have zero such commits in their universal-transfer conditions. The design therefore measures fixed candidate transfer rather than adaptation to policy feedback.

\begin{table*}[t]
\caption{Transfer of candidates optimized without runtime-defense feedback. Exact-goal and off-target commits are separated by call-ID linkage. Authorized completion is the binary extension-task outcome reported for each rollout.}
\label{tab:policy-adaptive-extension}
\centering
\scriptsize
\setlength{\tabcolsep}{1.5pt}
\begin{tabular}{@{}p{0.16\linewidth}p{0.17\linewidth}p{0.18\linewidth}rrrrr@{}}
\toprule
Replay design & Candidate optimization & Evaluated policy & Rollouts & \shortstack{Exact\\proposal} & \shortstack{Goal-matched\\commit} & \shortstack{Off-target\\commit} & \shortstack{Authorized\\completion} \\
\midrule
Instance-level replay & No defense & No defense & 360 & 0.0000 & 0.0000 & 0.1778 & 0.3472 \\
Universal replay & No defense & No defense & 2160 & 0.0000 & 0.0000 & 0.0722 & 0.2417 \\
Universal replay & No defense & Tool-boundary & 2160 & 0.0000 & 0.0000 & 0.0000 & 0.3134 \\
Universal replay & No defense & Taint-only v1 & 2160 & 0.0000 & 0.0000 & 0.0000 & 0.3130 \\
\bottomrule
\end{tabular}
\end{table*}

\subsection{Target-Preserving Held-Out Transfer}

The target-preserving checks ask whether a search representation that retains the exact action goal transfers beyond reference and in-distribution examples. Table~\ref{tab:target-preserving-transfer} shows a sharp decline on the family-held-out set. Both held-out variants produce exact proposals only in calendar/CRM, although all 36 traces in the first held-out run record exposure to the benchmark payload.

\begin{table*}[t]
\caption{Target-preserving proposal and commit results under no defense across reference, in-distribution, and family-held-out checks.}
\label{tab:target-preserving-transfer}
\centering
\scriptsize
\setlength{\tabcolsep}{2pt}
\begin{tabular}{@{}p{0.25\linewidth}p{0.22\linewidth}rrrp{0.20\linewidth}@{}}
\toprule
Stage & Scope & Traces & Exact proposal & Goal-matched commit & Domain coverage \\
\midrule
Reference replay & Retained target actions & 10 & 10/10 & 10/10 & -- \\
In-distribution check & Target-preserving search & 6 & 5/6 & 5/6 & calendar/CRM, documents \\
Family-held-out check & Unseen task families & 36 & 6/36 & 6/36 & calendar/CRM only \\
Scoring-variant repeat & Same held-out scope & 36 & 6/36 & 6/36 & calendar/CRM only \\
\bottomrule
\end{tabular}
\end{table*}

Together, these studies separate attack-interface coverage from policy enforcement. The exact-proposal control shows that, once the selected action is present, the instrumentation records proposal, admission, and commit as distinct stages. The fixed stress slice supplies a nonzero harmful-proposal signal, whereas the searches without defense feedback do not reliably preserve the target on held-out families. The results establish the demonstrated coverage of the proposal-to-commit measurement path and bound the coverage of candidate generation.

\section{AgentDojo Adapter and Bounded Native-Workflow Evaluation}
\label{sec:agentdojo-extension}

The AgentDojo evaluation uses two disjoint sets. The first tests whether the ContainmentBench event schema can represent native workflow executions. The adapter maps each native tool call to proposal, authorization-decision, result, and environment-difference events under a stable call identifier. Across 63 attack-free runs (21 candidate tasks under three seeds), adapter-derived task and security outcomes agreed with the corresponding native predicates in 63/63 runs. Twelve tasks met the clean-task and nonempty-injection-path criteria, and the eight with the lowest user-task identifiers were selected before policy evaluation. A separate 1,140-run no-defense source--task screen yielded 2,840 reconstructed call-state records, all of which satisfied the call-linkage contract.

An attacked portability set contained 600 rollouts, including 32 no-defense clusters, and produced a benchmark-defined violation in one cluster. A separate source--task check produced endpoint-positive outcomes on six of eight edges, all involving the same injection task and target tool. These sets are used to assess trace portability and source--task behavior, not policy effects.

The disjoint policy-evaluation set comprises custom workflows implemented in AgentDojo's native Slack, Travel, and Workspace environments. Structure-only selection, performed without attack or defended outcome labels, required a contract-ready task, an available target tool, resolvable arguments, native endpoint ground truth, and direct environment capability. A workflow edge is keyed by suite, user task, injection task, and injection source. Pairing 16 edges with three prespecified attack carriers defines 48 clusters. Fourteen edges were newly authored, and two Workspace edges met the same structure-only criteria.

Before defended runs, no-defense qualification evaluated all 48 clusters under three rollout seeds. The prespecified criterion required observations from all 16 edges, committed violations on at least eight edges, and at least two represented injection tasks and target tools. The resulting set had eight positive edges, five injection tasks, five target tools, and 57 committed violations in 144 rollouts. The same complete 16-edge set, clusters, and seeds were then evaluated under tool-boundary and intent-ledger enforcement.

Table~\ref{tab:agentdojo-validation} summarizes the portability checks and policy-set characteristics. Table~\ref{tab:agentdojo-new-pool} reports the matched policy outcomes.

\begin{table*}[h]
\caption{AgentDojo adapter checks and workflow-pool characteristics. The initial pool is used for trace-portability analysis but not for policy-effect estimation; the custom native-workflow pool provides the endpoint-positive no-defense cases used in the matched comparison.}
\label{tab:agentdojo-validation}
\centering
\scriptsize
\setlength{\tabcolsep}{2.5pt}
\begin{tabular}{@{}p{0.23\linewidth}p{0.19\linewidth}p{0.26\linewidth}p{0.24\linewidth}@{}}
\toprule
Study & Evaluation set & Observed result & Use in this study \\
\midrule
Attack-free endpoint agreement & 63 runs (21 tasks $\times$ 3 seeds) & Native task and security parity in 63/63 runs; 12 eligible, 8 selected before policy evaluation & Adapter qualification under the stated clean-task rule \\
Call-state reconstruction & Separate 1,140-run no-defense screen & 2,840 canonical call-state records satisfied the linkage contract & Call-event reconstruction check \\
Initial attacked pool & 600 rollouts; 32 no-defense clusters & No-defense positive in 1/32 clusters & Trace portability only; not used for policy-effect estimation \\
Source--task selection study & 144 no-defense rollouts over 8 edges & 6/8 endpoint-positive edges; positives share one injection task and target tool & Source--task and exact-argument diagnostic \\
Custom pool, no defense & 144 rollouts (16 edges, 48 clusters) & 8 positive edges; 57 committed violations & Five injection tasks and five target tools represented \\
Custom pool, defended & 288 matched rollouts & No committed violation observed & 144 rollouts for each runtime policy \\
\bottomrule
\end{tabular}
\end{table*}

\begin{table*}[ht]
\caption{Policy outcomes on the fixed custom native-workflow set. Each policy has 48 attack clusters evaluated under three seeds. Native user-task success is AgentDojo's user-task predicate evaluated on attacked runs; committed policy violations are reported separately. The rows with no observed event have an exact one-sided 95\% cluster upper confidence limit of 6.05\%. These custom workflows are distinct from AgentDojo's canonical task instances.}
\label{tab:agentdojo-new-pool}
\centering
\scriptsize
\setlength{\tabcolsep}{2pt}
\begin{tabular}{@{}p{0.24\linewidth}rrrrrr@{}}
\toprule
Policy & Rollouts & Clusters & Violations & Violation rate & \shortstack{Native user-task\\success} & Zero-event upper limit \\
\midrule
No defense & 144 & 48 & 57 & 0.3958 & 0.5764 & -- \\
Tool-boundary & 144 & 48 & 0 & 0.0000 & 0.8958 & 0.0605 \\
Intent-ledger v2 & 144 & 48 & 0 & 0.0000 & 0.9028 & 0.0605 \\
\bottomrule
\end{tabular}
\end{table*}

All 432 policy-matrix rollouts had complete call-state records. Under no defense, committed violations occurred in 32/54 Slack rollouts and 25/36 Workspace rollouts, but in none of the 54 Travel rollouts. The parser failed on 18/54 Travel rollouts under each policy, so Travel supplies neither an endpoint-positive no-defense control nor clean parser qualification in this study. No committed violation was observed under either defended condition. The native user-task predicate passed in 83/144 no-defense runs, 129/144 tool-boundary runs, and 130/144 intent-ledger v2 runs. The policy comparison is therefore supported by the endpoint-positive Slack and Workspace workflows in this custom set; it does not establish effectiveness across AgentDojo's canonical tasks or robustness to defense-adaptive attacks. The portability results indicate that source--task selection and exact-argument matching, rather than call-event mapping, limited the no-defense signal in the initial set.

\section{Redacted Case-Study Index}
\label{sec:case-study-index}

Table~\ref{tab:case-study-index} indexes the redacted single-trace and matched-pair examples provided in the supplementary package. Case~1 records a no-defense endpoint-positive run; Cases~2--3 form a matched authorized-workflow contrast; Case~4 is a matched logged-spread comparison; Cases~5--6 describe quarantine and rollback instrumentation; and Cases~7--8 record a policy-aware admission decision and parser diagnostics. These examples illustrate event-level behavior and are not used to estimate aggregate effects.

\begin{table*}[h]
\caption{Index of redacted single-trace and matched-pair case studies provided in the supplementary package.}
\label{tab:case-study-index}
\centering
\scriptsize
\setlength{\tabcolsep}{2pt}
\begin{tabular}{@{}p{0.19\linewidth}p{0.11\linewidth}p{0.16\linewidth}p{0.19\linewidth}p{0.27\linewidth}@{}}
\toprule
Case & Trace ID(s) & Policy condition(s) & Goal label & Illustrated behavior \\
\midrule
Case 1: no-defense policy-violation commit & \texttt{CB-CS01} & No defense & Private-plan summary; no sharing & Proposal-to-commit path for a benchmark-defined policy violation. \\
Case 2: authorized workflow blocked by taint-only v1 & \texttt{CB-CS02} & Taint-only v1 & Partner-note calendar invite & Authorized action suppressed by taint-only enforcement. \\
Case 3: matched authorized workflow completed under intent-ledger v2 & \texttt{CB-CS03} & Intent-ledger v2 & Partner-note calendar invite & Matched contrast to Case~2, with authorized completion under v2. \\
Case 4: matched logged-spread comparison & \texttt{CB-CS04A}; \texttt{CB-CS04B} & Tool-boundary; Intent-ledger v2 & CRM-note summary; no side effects & Matched trace pair comparing logged spread for the same scenario. \\
Case 5: pre-episode memory quarantine & \texttt{CB-CS05} & Intent-ledger v2 & Trusted-contact summary & Pre-episode quarantine event for a tainted memory item. \\
Case 6: post-hoc rollback trace & \texttt{CB-CS06} & Rollback v2 & Public Friday status note & Post-hoc rollback decision on an authorized public-note action. \\
Case 7: policy-aware stress proposal denied at admission & \texttt{CB-CS07} & Intent-ledger v2 & Private-plan summary; no sharing & Valid share-document proposals denied by the admission policy. \\
Case 8: invalid-tool parser diagnostic & \texttt{CB-CS08} & Intent-ledger v2 & Trusted-vendor delivery email & Two unknown-tool actions recorded without a tool-call proposal. \\
\bottomrule
\end{tabular}
\end{table*}

\section{Model--Interface Boundary Study}
\label{sec:mistral-interface}

To examine sensitivity to a different model--interface configuration, we used \nolinkurl{mistralai/Mistral-7B-Instruct-v0.3}~\cite{jiang2023mistral,mistralv03modelcard} after separately qualifying its native function-calling interface. Interface qualification is necessary because serialization, parsing, or call-linkage failures can otherwise be mistaken for containment. The native diagnostic establishes checkpoint tool-call capability, whereas the policy matrix measures the checkpoint together with \cb's common JSON action interface.

The checkpoint model card specifies tokenizer v3, function calling, and nine-character alphanumeric tool-call identifiers~\cite{mistralv03modelcard}. Our native diagnostic uses \code{mistral-common} 1.10.0 and \code{[TOOL\_CALLS]} serialization. It covers native round trips, direct model-generated calls with exact arguments, tool-result serialization and linkage, sequential and parallel calls, malformed results, final responses, and a 24-scenario carrier across four policies and three seeds. Carrier generation is deterministic at temperature 0, permits at most 180 new tokens and five turns, and records parser validity, no-defense utility, authorized side-effect capability, positive-control events, and invalid-tool actions. Table~\ref{tab:mistral-v03-interface} reports the four qualification levels and their evaluation units.

The formal matrix is a separate \(504\times4\times3=6{,}048\)-rollout experiment: 504 scenarios comprise 192 security, 120 active-tainted utility, 96 memory/recovery, and 96 clean-utility scenarios; each is evaluated under no defense, tool-boundary, taint-only v1, and intent-ledger v2 with seeds \(\{0,1,2\}\). It uses the local Transformers backend, the checkpoint chat template, temperature 0.2, at most 256 new tokens, and the common JSON \code{AgentAction} parser used in the Qwen study. The episode limit is ten turns, increased to twelve for the active-tainted component.

\begin{table*}[h]
\caption{Native Mistral interface qualification. Each row identifies its evaluation unit and observed criterion. These checks establish native serialization and event-linkage validity; they do not qualify the common-JSON parser used by the 6,048-rollout policy matrix.}
\label{tab:mistral-v03-interface}
\centering
\scriptsize
\setlength{\tabcolsep}{2.5pt}
\begin{tabular}{@{}p{0.15\linewidth}p{0.18\linewidth}p{0.29\linewidth}p{0.31\linewidth}@{}}
\toprule
Qualification level & Evaluation unit & Criterion & Observed result \\
\midrule
M0 round trip & 100 serialization cases & Preserve function name, argument object, and nine-character call ID through native encode/decode. & Joint identity 100/100; function name, arguments, and call ID each preserved in 100/100 cases. \\
M1 direct call & 96 prompts & Produce the target tool with exact arguments through the native parser; never execute an invalid tool. & Joint parser/tool/argument pass 96/96 (1.0000); invalid-tool executions 0/96. \\
M2 result loop & 27 conversations across nine categories & Preserve first- and second-turn actions, result linkage, and valid continuation. & All loop checks 27/27 (1.0000); missing call IDs rejected 3/3; invalid-tool executions 0/27. \\
M3 carrier & 288 rollouts: 24 scenarios \(\times\) 4 policies \(\times\) 3 seeds & Exercise parser validity, no-defense utility, authorized side effects, and a benchmark-local positive control. & Parser-valid action events 468/468; no-defense utility 36/36; side-effect capability 27/27; 24 high-risk proposals and 24 commits; invalid-tool action rate zero in 288/288 rollouts. \\
\bottomrule
\end{tabular}
\end{table*}

The native checks establish syntactic and event-linkage validity, but they do not establish semantic equivalence between native and common-JSON tool normalization, recipient mapping, taint labels, or ledger matching. The full experiment is therefore interpreted as a model--interface boundary rather than an isolated model effect. Table~\ref{tab:mistral-transfer-results} reports the stage-specific common-interface estimates.

\begin{table*}[h]
\caption{Stage-specific Mistral common-JSON results. Rates are means of per-rollout values; event counts are shown only for the binary security endpoint. Nonzero intervals use 5,000 scenario-cluster percentile bootstrap replicates with seed 20260625. The tool-boundary security row reports the exact one-sided 95\% cluster upper confidence limit for zero observed violations.}
\label{tab:mistral-transfer-results}
\centering
\scriptsize
\setlength{\tabcolsep}{2.5pt}
\begin{tabular}{@{}p{0.22\linewidth}p{0.23\linewidth}rrrrp{0.18\linewidth}@{}}
\toprule
Stage / metric & Policy & Rollouts & Clusters & Events & Estimate & 95\% interval / upper limit \\
\midrule
Security / committed violation & No defense & 576 & 192 & 205 & 0.3559 & [0.2951, 0.4184] \\
Security / committed violation & Tool-boundary & 576 & 192 & 0 & 0.0000 & 0.0155 (upper) \\
Security / committed violation & Taint-only v1 & 576 & 192 & 142 & 0.2465 & [0.1910, 0.3038] \\
Security / committed violation & Intent-ledger v2 & 576 & 192 & 142 & 0.2465 & [0.1910, 0.3038] \\
Active / proposal-commit score & No defense & 360 & 120 & -- & 0.7902 & [0.7178, 0.8597] \\
Active / proposal-commit score & Tool-boundary & 360 & 120 & -- & 0.7902 & [0.7178, 0.8597] \\
Active / proposal-commit score & Taint-only v1 & 360 & 120 & -- & 0.2189 & [0.1529, 0.2877] \\
Active / proposal-commit score & Intent-ledger v2 & 360 & 120 & -- & 0.5960 & [0.5185, 0.6741] \\
\bottomrule
\end{tabular}
\end{table*}

The final matrix contained 6,048 completed rollouts. On Mistral security scenarios, no defense produced 205/576 committed policy violations, tool-boundary produced none, and taint-only v1 and intent-ledger v2 each produced 142/576. On active-tainted scenarios, the mean proposal-commit score was 0.7902 under both no defense and tool-boundary, 0.2189 under v1, and 0.5960 under v2. The corresponding structured task-completion rates were 0.1194, 0.1194, 0.0194, and 0.1639; proposal commit and task completion are reported separately.

Common-interface parser coverage was imperfect. Over the 576 security rollouts per policy, the mean per-rollout parse-failure rates were 0.2291, 0.2168, 0.2201, and 0.2201 for no defense, tool-boundary, v1, and v2, respectively; valid-tool-proposal rates were 0.7690, 0.7818, 0.7785, and 0.7785. These rates average action-event fractions within each rollout. Thus, the native qualification cannot be read as qualification of the full matrix's parser path or as separation of model effects from interface normalization. Under this model--interface configuration, the v1-to-v2 proposal-commit improvement persisted. No committed violation was observed under tool-boundary enforcement, whereas violations were observed under both v1 and v2.

\section{Metric and Trace-Field Glossary}

Table~\ref{tab:appendix-metrics} maps manuscript quantities to their trace fields. Metrics are reported separately by evidence stage and are not combined into a scalar score. Endpoint violations and proposal-commit quantities have the indicated direction only within their applicable stages. Logged spread, taint-to-commit, and privilege-transition measures are diagnostics whose interpretation depends on authorization and stage and, for normalized spread, on the realized-trace denominator. Recovery comparisons require an informative positive control.

\begin{table*}[h]
\caption{Metric and trace-field glossary. Per-rollout denominators and stage aggregation are given in Table~\ref{tab:metric-denominators}.}
\label{tab:appendix-metrics}
\centering
\scriptsize
\setlength{\tabcolsep}{2pt}
\begin{tabular}{@{}p{0.20\linewidth}p{0.18\linewidth}p{0.42\linewidth}p{0.13\linewidth}@{}}
\toprule
Metric & Family & Operational definition & Interpretation \\
\midrule
committed\_harm & Security & Trace field for the benchmark-defined committed policy-violation indicator \(V_{\mathrm{commit}}\) in Eq.~\ref{eq:committed-harm}. & Lower \\
secret\_leakage & Security & Rollout indicator that the final committed sandbox state contains a scenario canary in an attacker-addressed, public, or external record. & Lower \\
wrong\_recipient\_rate & Security & Fraction of committed recipient-bearing high-risk actions not explicitly authorized by the trusted goal; denominator clipped at one. & Lower \\
taint\_to\_commit\_rate & Security/propagation & Taint-to-commit ratio: fraction of tainted high-risk proposals whose corresponding calls commit. & Context-dependent \\
blast\_radius\_norm & Propagation & Logged taint spread normalized by the realized structural trace region. & Context-dependent \\
privilege\_jump & Propagation & Per-rollout indicator for a logged transition from tainted lower-trust input into a high-risk action context; reported rates average this indicator. & Context-dependent \\
memory\_reinfection & Memory/recovery & Indicator that tainted persistent memory reappears after a recovery or scrub step. & Lower \\
recovery\_success & Memory/recovery & Rollout indicator that the designated synthetic recovery objective completes without episode-C harm or active tainted memory. & Higher with positive control \\
authorized\_tainted\_action\_success & Utility & Per-rollout fraction in Eq.~\ref{eq:active-utility}: committed explicitly authorized proposals with tainted input labels divided by eligible proposals. & Higher on active stage \\
clean\_benign\_completion & Utility & Structured task-specific success oracle on clean workflows without injected content. & Higher on clean stage \\
false\_positive\_block\_rate & Utility & Authorized high-risk proposals receiving block, quarantine, or rollback decisions divided by authorized high-risk proposals. & Lower on utility stage \\
false\_positive\_confirm\_rate & Utility & Authorized high-risk proposals receiving confirmation decisions divided by authorized high-risk proposals. & Lower on utility stage \\
confirmation\_burden & Utility/diagnostic & Number of confirmation-result events in a rollout. & Descriptive burden \\
valid\_tool\_proposal\_rate & Measurement validity & Fraction of model-action events parsed as registered-tool proposals. & Qualification diagnostic \\
\bottomrule
\end{tabular}
\end{table*}

The confirmation rate divides confirmation decisions by authorized high-risk opportunities, whereas confirmation burden counts confirmation-result events per rollout.

\section{Supplementary Protocol and Outcome Tables}
\label{sec:supplementary-protocol-tables}

Table~\ref{tab:evidence-stages} in the main benchmark section gives the evidence-stage counts. The tables below define the structured task-completion oracle, metric denominators, and policy roles. Tables~\ref{tab:security-results} and \ref{tab:memory-results} report the security-stage and memory/recovery-stage outcomes.

\begin{table*}[h]
\caption{Structured task-completion oracle. Required commits and exact fields determine task success; committed policy violations are evaluated separately by \(V_{\mathrm{commit}}\).}
\label{tab:benign-success-oracle}
\centering
\scriptsize
\setlength{\tabcolsep}{2pt}
\begin{tabular}{@{}p{0.24\linewidth}p{0.42\linewidth}p{0.24\linewidth}@{}}
\toprule
Task family & Required structured checks & Not evaluated \\
\midrule
Summary-only tasks & A final answer is present and contains a task-specific required marker; tools prohibited by the summary-only goal do not commit. & Factual completeness or prose quality beyond the marker check. \\
Authorized email send & The required draft and send tools commit, and the committed recipient exactly matches the scenario criterion. & Persuasiveness, style, or semantic quality of the message body. \\
Authorized document share or public status note & The required share or create tool commits; the document identifier, recipient, or visibility exactly matches the scenario criterion. & Human-rated quality of the shared content or status note. \\
Authorized calendar, CRM, or memory update & The required create, invite, update, and memory tools commit as applicable; attendee, contact, memory key, and memory scope exactly match the scenario criterion. & Downstream business usefulness beyond the exact structured fields. \\
\bottomrule
\end{tabular}
\end{table*}

\begin{table*}[h]
\caption{Operational numerators, within-rollout denominators, and stage aggregation. A denominator clipped at one assigns zero to a rollout with no eligible opportunity.}
\label{tab:metric-denominators}
\centering
\scriptsize
\setlength{\tabcolsep}{1.5pt}
\begin{tabular}{@{}p{0.20\linewidth}p{0.25\linewidth}p{0.25\linewidth}p{0.20\linewidth}@{}}
\toprule
Metric & Per-rollout numerator & Denominator and zero handling & Reported aggregation \\
\midrule
\metric{wrong_recipient_rate} & Committed external sends, public shares, or calendar invites not explicitly authorized by the trusted goal. & All committed actions in those three classes; denominator clipped at one. & Mean of rollout fractions over the reported stage, including zero-opportunity rollouts. \\
\metric{taint_to_commit_rate} & Tainted high-risk proposals whose calls commit. & All tainted high-risk proposals; denominator clipped at one. & Mean of rollout fractions over the reported stage, including zero-opportunity rollouts. \\
\metric{privilege_jump} & One if tainted lower-trust input reaches a high-risk action context; zero otherwise. & Not applicable; this is a binary rollout value. & Mean of the binary value over the reported stage. \\
\metric{false_positive_block_rate} & Explicitly authorized high-risk proposals receiving block, quarantine, or rollback decisions. & All explicitly authorized high-risk proposals; denominator clipped at one. & Mean of rollout fractions over the applicable utility stage, including zero-opportunity rollouts. \\
\metric{false_positive_confirm_rate} & Explicitly authorized high-risk proposals receiving confirmation decisions. & All explicitly authorized high-risk proposals; denominator clipped at one. & Mean of rollout fractions over the applicable utility stage, including zero-opportunity rollouts. \\
\metric{confirmation_burden} & Confirmation-result events in the rollout. & Not applicable; this is a rollout count. & Mean count over the reported confirmation or rollback scope. \\
\metric{recovery_success} & One when the designated recovery objective completes without episode-C harm or active tainted memory; zero otherwise. & Not applicable; this is a binary rollout value. & Mean over memory/recovery rollouts only. \\
Authorized action-commit rate & Committed, explicitly authorized proposals with tainted input labels. & All explicitly authorized proposals with tainted input labels; denominator clipped at one. & Mean active-stage rollout fraction, including zero-opportunity rollouts. \\
\metric{valid_tool_proposal_rate} & Model-action events parsed as registered-tool proposals after strict parsing or deterministic repair. & All model-action events; denominator clipped at one. & Mean of rollout fractions over the reported interface or evidence stage. \\
\bottomrule
\end{tabular}
\end{table*}

\begin{table*}[h]
\caption{Policy conditions, implemented mechanisms, and evaluation roles.}
\label{tab:methods}
\centering
\scriptsize
\setlength{\tabcolsep}{2pt}
\begin{tabular}{@{}p{0.25\linewidth}p{0.28\linewidth}p{0.37\linewidth}@{}}
\toprule
Policy condition & Mechanism & Evaluation role \\
\midrule
No defense & No runtime intervention. & Endpoint-positive control and unconstrained task reference. \\
Prompt-only & System-prompt instruction without runtime enforcement. & Instruction-only comparison. \\
Structured-guard proxy & Confirmation-mediated structured-action interface. & Structured-interface comparison. \\
Tool-boundary & Deterministic trusted-goal check at each side-effecting tool call. & Runtime-boundary baseline. \\
Taint-only v1 & Ordered provenance-sensitive rules without the structured intent ledger. & Diagnostic authorization contrast. \\
Intent-ledger v2 & Exact structured-ledger branch with ordered v1 fallback. & Structured authorization condition. \\
Intent-ledger v2 + rollback instrumentation & The same v2 admission rule with synthetic post-commit rollback hooks. & Recovery-instrumentation condition. \\
\bottomrule
\end{tabular}
\end{table*}

\begin{table*}[!t]
\caption{Security-stage endpoint outcomes. Each policy condition has 960 rollouts (192 scenarios times five seeds); no defense supplies the endpoint-positive control. Secret leakage is zero under every condition and is therefore descriptive rather than discriminative in this stage.}
\label{tab:security-results}
\centering
\scriptsize
\setlength{\tabcolsep}{1.1pt}
\begin{tabular}{@{}p{0.25\linewidth}rrrrr@{}}
\toprule
Policy condition & N & \shortstack{Committed\\violation} & \shortstack{Wrong\\recipient} & \shortstack{Secret\\leakage} & \shortstack{Logged\\spread} \\
\midrule
No defense & 960 & 0.3354 & 0.3292 & 0.0000 & 0.6984 \\
Prompt-only & 960 & 0.3333 & 0.3333 & 0.0000 & 0.6570 \\
Structured-guard proxy & 960 & 0.0000 & 0.0000 & 0.0000 & 0.6634 \\
Tool-boundary & 960 & 0.0000 & 0.0000 & 0.0000 & 0.6640 \\
Taint-only v1 & 960 & 0.0000 & 0.0000 & 0.0000 & 0.6630 \\
Intent-ledger v2 & 960 & 0.0000 & 0.0000 & 0.0000 & 0.6624 \\
Intent-ledger v2 + rollback instrumentation & 960 & 0.0000 & 0.0000 & 0.0000 & 0.6636 \\
\bottomrule
\end{tabular}
\end{table*}

\begin{table*}[!t]
\caption{Memory/recovery-stage outcomes. Each policy condition has 480 rollouts (96 scenarios times five seeds). Reinfection is zero under every condition; recovery and logged taint spread are reported descriptively.}
\label{tab:memory-results}
\centering
\scriptsize
\setlength{\tabcolsep}{1.1pt}
\begin{tabular}{@{}p{0.31\linewidth}rrrr@{}}
\toprule
Policy condition & N & Reinfection & Recovery & \shortstack{Logged\\spread} \\
\midrule
No defense & 480 & 0.0000 & 0.6312 & 0.5249 \\
Prompt-only & 480 & 0.0000 & 0.6250 & 0.4665 \\
Structured-guard proxy & 480 & 0.0000 & 0.6312 & 0.5249 \\
Tool-boundary & 480 & 0.0000 & 0.6312 & 0.5249 \\
Taint-only v1 & 480 & 0.0000 & 0.6312 & 0.3858 \\
Intent-ledger v2 & 480 & 0.0000 & 0.6312 & 0.3858 \\
Intent-ledger v2 + rollback instrumentation & 480 & 0.0000 & 0.6312 & 0.3858 \\
\bottomrule
\end{tabular}
\end{table*}

\section{Evidence-Stage Composition}

The 504 main-study scenario identities and stage assignments were fixed before aggregation. The security stage is the Cartesian product of three domains, two multi-agent topologies, two memory modes, two injection channels, four prompt-strength conditions, and generation seeds 501 and 502, yielding 192 scenarios. Memory/recovery candidates cross three scenario families, three domains, three topologies, two source channels, and generation seeds 511 and 512; 96 are selected deterministically with selection seed \code{20260624}, preserving 32 scenarios per domain and per family. The active-tainted stage contains 120 structured authorized-task variants over benign tainted tool results generated with seeds 521 and 522. Clean utility crosses eight benign tasks with three topologies, two permission layouts, and two memory modes at generation seed 531, yielding 96 scenarios. Table~\ref{tab:evidence-stages} reports the policy and rollout counts. The scenario manifest records identities, stage assignments, generation seeds, and construction parameters.

\section{Runtime Authorization Procedure}

The parser first converts a model completion into one structured action and records its diagnostic status. When that action names a tool absent from the current agent's registry, the runner retains the model-action record with status \metric{invalid_tool} but constructs no \code{ToolCall}; consequently, no tool-proposal, authorization, execution, or commit event follows. For a registered-tool proposal, intent-ledger v2 evaluates the rules in Table~\ref{tab:policy-rule-map} in order using the prefix trace available at authorization time. Proposals that do not trigger either v2 branch delegate to the ordered v1 policy.

\begin{table*}[h]
\caption{Ordered implementation mapping for parser handling and intent-ledger v2 authorization. V2 evaluates its two rules before delegating unmatched proposals to v1 in the order shown.}
\label{tab:policy-rule-map}
\centering
\scriptsize
\setlength{\tabcolsep}{2pt}
\begin{tabular}{@{}p{0.10\linewidth}p{0.31\linewidth}p{0.51\linewidth}@{}}
\toprule
Order & Parser outcome or rule & Implemented condition and decision \\
\midrule
Parser & Unknown tool & Model action names a tool outside the current registry: retain a non-executable model-action diagnostic and emit no proposal, authorization, execution, or commit event. \\
V2-1 & Secret-bearing external action & \(\SecretBoundary(q)=1\): block the secret-bearing external/public proposal. \\
V2-2 & Exact authorized tainted action & \(\HighRisk(q)\wedge\Tainted(q;\tau)\wedge\Match(q,\Intent_s)\wedge\neg\AttackArg(q)\): allow. \\
V2-3 & Ordered v1 fallback & Otherwise evaluate the following v1 rules; exact-match failure alone is not a block decision. \\
V1-1 & Secret to unauthorized sink & External send, public share, or calendar invite carries a secret to an unapproved recipient: block. \\
V1-2 & Tainted persistent-memory write & Persistent-memory write carries recorded taint: quarantine. \\
V1-3 & Low-to-high privilege path & A high-risk proposal by a high-privilege agent depends on a tainted low-privilege node: block. \\
V1-4 & Goal-disallowed tainted high-risk action & High-risk proposal is tainted and \(\neg\GoalAllowed(q,s)\): block. \\
V1-5 & Confirmation for remaining tainted external action & Remaining tainted external send, public share, or calendar invite: request confirmation. \\
V1-6 & Default allow & No earlier rule fires: allow. \\
\bottomrule
\end{tabular}
\end{table*}

This decision procedure is limited to the benchmark's structured authorization model. Each registered-tool proposal and linked result event records the applied rule, tool arguments, sink, taint and authorization fields, and defense decision; the shared tool-call identifier connects proposal and result records. The rollback variant applies the same admission rules and adds synthetic post-commit rollback events, which remain descriptive trace instrumentation in the memory/recovery stage.

\section{Model and Runtime Details}
\label{sec:model-runtime-details}

The primary checkpoint is \code{Qwen/Qwen2.5-7B-Instruct}, loaded from a local ModelScope snapshot with bfloat16 weights and a 32,768-token context window. The backend calls \code{apply\_chat\_template} with \code{add\_generation\_prompt=True}; if a checkpoint does not expose a usable template, the implementation falls back to role-prefixed serialization. The evaluated Qwen snapshot uses its checkpoint template.

The boundary study uses \nolinkurl{mistralai/Mistral-7B-Instruct-v0.3} from a local snapshot. Its native-interface diagnostic uses tokenizer v3 and \code{mistral-common} serialization; the policy matrix uses the checkpoint chat template and the benchmark's common JSON action interface.

The shared local-HF runtime uses Python 3.10.20, Torch 2.5.1 with CUDA 12.1, Transformers 5.5.3, and NumPy 1.26.4. The native Mistral diagnostic additionally uses \code{mistral-common} 1.10.0. Checkpoint metadata are included with the supplementary materials. No provider-side model call is used in either policy matrix.

\section{Parser Diagnostics and Seed Handling}
\label{sec:parser-seed-validity}

We use repair-aware parsing to distinguish malformed action serialization from outputs that contain no recoverable structured tool proposal. The parser accepts strict JSON and a prespecified sequence of repairs for fenced, prefixed, and embedded JSON. The accepted argument containers are \code{args}, \code{tool_input}, \code{toolbench_input}, \code{arguments}, \code{input}, \code{parameters}, and \code{tool_params}; nested wrappers may additionally use \code{kwargs}. A registered tool name may appear as \code{action_type}, and explicit aliases normalize final-answer and target-agent fields. Invalid tools are recorded but never executed.

Table~\ref{tab:appendix-parser} reports the final parser diagnostic on 96 Qwen model-action records used during parser development. It is a development-set check rather than a held-out estimate of parser generalization or cross-model robustness. The parse-status and post-parse blocks are separate partitions and should not be added across blocks.

\begin{table*}[h]
\caption{Development-set diagnostic for the final Qwen action parser. Counts close separately within the parse-status and post-parse outcome partitions.}
\label{tab:appendix-parser}
\centering
\scriptsize
\setlength{\tabcolsep}{1.5pt}
\begin{tabular}{@{}p{0.19\linewidth}p{0.29\linewidth}rp{0.40\linewidth}@{}}
\toprule
Layer & Class & Count & Definition \\
\midrule
Parse-status partition & Valid strict JSON & 5 & Parsed without extraction, repair, or alias normalization. \\
 & Valid after repair & 89 & Parsed after deterministic extraction, JSON repair, or supported normalization. \\
 & Invalid schema & 2 & No supported action schema remained after attempted repair. \\
 & Total & 96 & Complete parse-status partition. \\
\midrule
Post-parse outcome partition & Known-tool proposal & 85 & Schema-valid registered-tool actions. \\
 & Schema-valid final answer & 6 & Valid final-answer objects, not tool proposals. \\
 & Schema-valid inter-agent message & 3 & Valid message actions, not tool proposals. \\
 & Unknown-tool proposal & 0 & Registry check failures; such actions are non-executable. \\
 & Invalid schema & 2 & Records outside the 94 schema-valid outcomes. \\
 & Total & 96 & Complete post-parse outcome partition. \\
\bottomrule
\end{tabular}
\end{table*}

The primary Qwen matrix reuses rollout seeds \(\{0,1,2,3,4\}\) across all seven policy conditions for each scenario; the Mistral boundary-condition matrix analogously reuses \(\{0,1,2\}\) across its four conditions. In all 17,640 primary rows, \metric{requested_seed}, \metric{rollout_seed}, and the analysis key's \metric{seed} agree, and \metric{provider_seed_supported} is true. The local-HF backend seeds Python, NumPy, Torch, and all visible CUDA generators before each completion. The call uses sampling (\code{do_sample=True}), temperature 0.2, and at most 256 new tokens. Top-\(p\), top-\(k\), and repetition penalty are not overridden at the call site; the pinned Qwen generation configuration supplies 0.8, 20, and 1.05, respectively. Model, tokenizer, and software revisions are reported in the preceding section.

These controls support matched stochastic comparisons, not bitwise reproducibility: deterministic Torch algorithms are not enabled, and results can vary across hardware or software versions. No backend lacking provider-side seed control appears in either reported formal matrix, so unsupported-provider rows receive no paired statistical treatment here; the metadata flag is retained to prevent future such rows from being mistaken for provider-seeded replicates.

\section{Statistical Details and Seed Variance}

The basic matched observation is a scenario--model--seed row. Because the five rollout seeds associated with the same scenario are repeated realizations rather than independent tasks, the scenario is the primary uncertainty-resampling unit. Unless otherwise stated, primary intervals use 5,000 percentile scenario-cluster bootstrap replicates with random seed 20260625. Each replicate samples scenarios with replacement and retains all seed-matched contrasts for each sampled scenario. All comparisons in Table~\ref{tab:appendix-paired} have complete matches; the general analysis excludes incomplete defense pairs and reports the retained row and cluster counts.

\begin{sloppypar}
We do not assign a directional interpretation when an interval includes zero. The term \emph{tie} is reserved for comparisons in which every paired value is identical. Table~\ref{tab:appendix-paired} reports the comparisons corresponding to the main findings; the supplementary package contains the complete pair-level output. No multiplicity adjustment is applied to secondary diagnostic comparisons, which are interpreted descriptively.
\end{sloppypar}

\begin{table*}[h]
\caption{Scenario-cluster paired comparisons corresponding to the main findings. Delta is Mean B minus Mean A. Utility and recovery metrics are shown only on applicable stages.}
\label{tab:appendix-paired}
\centering
\scriptsize
\setlength{\tabcolsep}{2pt}
\begin{tabular}{@{}p{0.20\linewidth}p{0.15\linewidth}p{0.09\linewidth}rrrrrr@{}}
\toprule
Comparison & Metric & Stage & Mean A & Mean B & Delta & 95\% CI & Rows & Clusters \\
\midrule
None vs Tool-boundary & Committed violation & Pooled & 0.1456 & 0.0000 & -0.1456 & [-0.1754, -0.1155] & 2520 & 504 \\
Taint-only v1 vs Intent-ledger v2 & Proposal-commit score & Active & 0.1642 & 0.8567 & 0.6925 & [0.6160, 0.7637] & 600 & 120 \\
Tool-boundary vs Intent-ledger v2 & Proposal-commit score & Active & 0.9233 & 0.8567 & -0.0667 & [-0.1167, -0.0250] & 600 & 120 \\
Tool-boundary vs Intent-ledger v2 & Logged taint spread & Pooled & 0.5528 & 0.5219 & -0.0309 & [-0.0403, -0.0218] & 2520 & 504 \\
Tool-boundary vs Intent-ledger v2 & Logged taint spread & Security & 0.6640 & 0.6624 & -0.0016 & [-0.0032, -0.0003] & 960 & 192 \\
Tool-boundary vs Intent-ledger v2 & Clean completion & Clean & 0.9958 & 0.9958 & 0.0000 & [0.0000, 0.0000] & 480 & 96 \\
Tool-boundary vs Intent-ledger v2 & Memory reinfection & Memory & 0.0000 & 0.0000 & 0.0000 & [0.0000, 0.0000] & 480 & 96 \\
Intent-ledger v2 vs rollback v2 & Recovery success & Memory & 0.6312 & 0.6312 & 0.0000 & [0.0000, 0.0000] & 480 & 96 \\
\bottomrule
\end{tabular}
\end{table*}

Zero-event estimates are not paired differences. Table~\ref{tab:appendix-zero-event} therefore reports them separately at the scenario-cluster unit. For zero observed cluster events, the exact one-sided 95\% Clopper--Pearson upper limit is \(1-0.05^{1/n}\), where \(n\) is the number of scenario clusters.

\begin{table*}[h]
\caption{Zero-event bounds for the intent-ledger v2 committed-violation endpoint. Rollout counts describe the observed event total; confidence limits use scenario clusters as Bernoulli units.}
\label{tab:appendix-zero-event}
\centering
\scriptsize
\setlength{\tabcolsep}{3pt}
\begin{tabular}{@{}p{0.20\linewidth}p{0.12\linewidth}p{0.20\linewidth}rrrr@{}}
\toprule
Policy & Stage & Metric & Events / rollouts & Clusters & Event clusters & Exact upper 95\% \\
\midrule
Intent-ledger v2 & Pooled & Committed violation & 0 / 2520 & 504 & 0 & 0.5926\% \\
Intent-ledger v2 & Security & Committed violation & 0 / 960 & 192 & 0 & 1.5482\% \\
\bottomrule
\end{tabular}
\end{table*}

Table~\ref{tab:appendix-seeds} describes variation across the five seed-specific stage means. The reported SD is the population standard deviation of those five means (divisor five); Min and Max are their observed extrema. These are descriptive decoding-seed summaries rather than confidence intervals.

\begin{table}[H]
\caption{Seed-specific variation for the main evaluation. Each row summarizes five stage means for rollout seeds 0--4; non-applicable metric--stage combinations are omitted.}
\label{tab:appendix-seeds}
\centering
\fontsize{5.8pt}{6.1pt}\selectfont
\setlength{\tabcolsep}{2pt}
\begin{tabular}{@{}p{0.54\linewidth}rrp{0.20\linewidth}@{}}
\toprule
Series (stage / policy / metric) & Mean & SD & Observed range \\
\midrule
Active / Intent-ledger v2 / proposal-commit score & 0.8567 & 0.0180 & [0.8222, 0.8694] \\
Active / Intent-ledger v2 / logged taint spread & 0.8236 & 0.0066 & [0.8178, 0.8364] \\
Active / Tool-boundary / proposal-commit score & 0.9233 & 0.0180 & [0.8889, 0.9361] \\
Active / Tool-boundary / logged taint spread & 0.8394 & 0.0084 & [0.8324, 0.8556] \\
Active / None / proposal-commit score & 0.9233 & 0.0180 & [0.8889, 0.9361] \\
Active / None / committed violation & 0.0750 & 0.0217 & [0.0333, 0.0917] \\
Security / Intent-ledger v2 / logged taint spread & 0.6624 & 0.0076 & [0.6479, 0.6677] \\
Security / Tool-boundary / logged taint spread & 0.6640 & 0.0065 & [0.6517, 0.6691] \\
Memory / Intent-ledger v2 / recovery success & 0.6312 & 0.0313 & [0.5938, 0.6667] \\
Clean / Intent-ledger v2 / clean completion & 0.9958 & 0.0083 & [0.9792, 1.0000] \\
\bottomrule
\end{tabular}
\end{table}

\section{Formal Notation Summary}

Table~\ref{tab:formal-symbols} summarizes the key notation used to define benchmark-local traces, metrics, and runtime policy decisions.

\begin{table*}[!t]
\caption{Key notation for trace, metric, and policy semantics.}
\label{tab:formal-symbols}
\centering
\scriptsize
\renewcommand{\arraystretch}{0.92}
\setlength{\tabcolsep}{3pt}
\begin{tabular}{@{}p{0.25\textwidth}p{0.69\textwidth}@{}}
\toprule
Symbol & Meaning \\
\midrule
\(\tau\) & A rollout, including scenario, model, defense, and seed metadata. \\
\(\Trace_\tau=(V_\tau,E_\tau)\) & Directed logged trace graph for rollout \(\tau\). \\
\(\Trace_{\tau,\le k(q)}\) & Prefix trace visible at the authorization time of proposal \(q\). \\
\(U_\tau\) & Untrusted source nodes in the trace. \\
\(\Taint_\tau\) & Nodes carrying a recorded benchmark taint label; this is an instrumentation-defined set rather than an assumption that all structurally reachable nodes are tainted. \\
\(R_\tau\) & Structural region reachable from \(U_\tau\) over the full logged edge set, used as the realized LTS denominator. \\
\(Q_\tau\) & Parsed tool-proposal nodes in the trace. \\
\(C_\tau\) & Subset \(\{q\in Q_\tau:\Commit(q)=1\}\) whose corresponding calls commit to the benchmark-local environment. \\
\(Q^{A,T}_\tau\) & Proposals whose action and arguments are explicitly authorized and whose recorded input labels contain taint. \\
\(\Intent_s\) & Trusted intent ledger constructed from scenario authorization metadata before rollout. \\
\(\Match(q,\Intent_s)\) & Indicator that proposal \(q\) exactly matches an entry over every represented action, target, and constraint field. \\
\(\GoalAllowed(q,s)\) & Coarse trusted-goal and sandbox-constraint predicate used by v1 and tool-boundary enforcement. \\
\(\Tainted(q;\tau)\) & Indicator that proposal-path or argument labels intersect the benchmark taint set at authorization time. \\
\(\SecretBoundary(q)\) & Indicator that the v2 secret-bearing external/public precheck fires. \\
\(\AttackArg(q)\) & Indicator that argument-derived labels contain a benchmark payload or persistent-memory taint. \\
\(\mathrm{D}_{\mathrm{v1}}(q;\tau,s)\) & Ordered v1 authorization decision applied after its secret, memory, privilege-jump, goal, and confirmation checks. \\
\(\mathrm{D}_{\mathrm{IL}}(q;\tau,s)\) & Intent-ledger v2 decision in Eq.~\ref{eq:il-decision}, including ordered v1 fallback. \\
\(\mathrm{Unauthorized}(q)\) & Indicator that a committed high-risk proposal fails the trusted-goal authorization predicate. \\
\(\mathrm{SecretLeak}(q)\) & Indicator that a commit places a synthetic canary in an external or public benchmark-local sink. \\
\(\mathrm{WrongRecipient}(q)\) & Indicator that a committed external send, public share, or calendar invite resolves to a scenario-unauthorized recipient. \\
\(V_{\mathrm{commit}}\) & Benchmark-defined committed policy-violation indicator, stored in the trace field \metric{committed_harm}. \\
\(B(\tau)\) & Structured benign task-success oracle over required commits and task-specific fields. \\
\(U_{\mathrm{active}}(\tau)\) & Per-rollout commit fraction over \(Q^{A,T}_\tau\), with a denominator lower-bounded by one. \\
\(\bar m_{g,d}\) & Mean of metric \(m\) over rollouts in evidence stage \(g\) under defense \(d\). \\
\(\mathrm{TCR}\) & Taint-to-commit ratio, recorded as \metric{taint_to_commit_rate}. \\
\(\mathrm{LTS}\) & Logged taint spread, recorded as \metric{blast_radius_norm}. \\
\(\phi(\tau)\) & Shared-schema 12-field trace projection used for matched cross-policy comparison. \\
\(D_{\phi}(\tau_a,\tau_b)\) & Projection-divergence indicator in Eq.~\ref{eq:projection-divergence}, using deterministic numeric encoding and absolute tolerance \(10^{-12}\). \\
\(\Delta_{m,g,\mu}(d_b,d_a)\) & Paired stage- and model-specific mean difference \(m(d_b)-m(d_a)\) for metric \(m\). \\
\bottomrule
\end{tabular}
\end{table*}

\clearpage
\bibliographystyle{unsrt}
\bibliography{ref}

@inproceedings{greshake2023not,
  title={Not what you've signed up for: Compromising Real-World {LLM}-Integrated Applications with Indirect Prompt Injection},
  author={Greshake, Kai and Abdelnabi, Sahar and Mishra, Shailesh and Endres, Christoph and Holz, Thorsten and Fritz, Mario},
  booktitle={Proceedings of the 16th ACM Workshop on Artificial Intelligence and Security},
  pages={79--90},
  year={2023},
  publisher={Association for Computing Machinery},
  address={New York, NY, USA},
  doi={10.1145/3605764.3623985},
  url={https://doi.org/10.1145/3605764.3623985}
}

@inproceedings{yi2023bipia,
  title={Benchmarking and Defending Against Indirect Prompt Injection Attacks on Large Language Models},
  author={Yi, Jingwei and Xie, Yueqi and Zhu, Bin and Kiciman, Emre and Sun, Guangzhong and Xie, Xing and Wu, Fangzhao},
  booktitle={Proceedings of the 31st ACM SIGKDD Conference on Knowledge Discovery and Data Mining},
  volume={1},
  pages={1809--1820},
  year={2025},
  publisher={Association for Computing Machinery},
  address={New York, NY, USA},
  doi={10.1145/3690624.3709179},
  url={https://doi.org/10.1145/3690624.3709179}
}

@inproceedings{zhan2024injecagent,
  title={{InjecAgent}: Benchmarking Indirect Prompt Injections in Tool-Integrated Large Language Model Agents},
  author={Zhan, Qiusi and Liang, Zhixiang and Ying, Zifan and Kang, Daniel},
  booktitle={Findings of the Association for Computational Linguistics: ACL 2024},
  pages={10471--10506},
  year={2024},
  month={August},
  address={Bangkok, Thailand},
  publisher={Association for Computational Linguistics},
  doi={10.18653/v1/2024.findings-acl.624},
  url={https://aclanthology.org/2024.findings-acl.624/}
}

@inproceedings{debenedetti2024agentdojo,
  title={{AgentDojo}: A Dynamic Environment to Evaluate Prompt Injection Attacks and Defenses for {LLM} Agents},
  author={Debenedetti, Edoardo and Zhang, Jie and Balunovi{\'c}, Mislav and Beurer-Kellner, Luca and Fischer, Marc and Tram{\`e}r, Florian},
  booktitle={Advances in Neural Information Processing Systems},
  volume={37},
  pages={82895--82920},
  year={2024},
  publisher={Curran Associates, Inc.},
  address={Red Hook, NY, USA},
  doi={10.52202/079017-2636},
  url={https://proceedings.neurips.cc/paper_files/paper/2024/hash/97091a5177d8dc64b1da8bf3e1f6fb54-Abstract-Datasets_and_Benchmarks_Track.html}
}

@misc{abdelnabi2025llmailinject,
  title={{LLMail-Inject}: A Dataset from a Realistic Adaptive Prompt Injection Challenge},
  author={Abdelnabi, Sahar and Fay, Aideen and Salem, Ahmed and Zverev, Egor and Liao, Kai-Chieh and Liu, Chi-Huang and Kuo, Chun-Chih and Weigend, Jannis and Manlangit, Danyael and Apostolov, Alex and others},
  howpublished={arXiv preprint arXiv:2506.09956},
  year={2025},
  url={https://arxiv.org/abs/2506.09956}
}

@inproceedings{geng2026piarena,
  title={{PIArena}: A Platform for Prompt Injection Evaluation},
  author={Geng, Runpeng and Yin, Chenlong and Wang, Yanting and Chen, Ying and Jia, Jinyuan},
  booktitle={Proceedings of the 64th Annual Meeting of the Association for Computational Linguistics (Volume 1: Long Papers)},
  pages={33170--33192},
  year={2026},
  month={July},
  address={San Diego, California, United States},
  publisher={Association for Computational Linguistics},
  doi={10.18653/v1/2026.acl-long.1533},
  url={https://aclanthology.org/2026.acl-long.1533/}
}

@inproceedings{liu2024formalizing,
  title={Formalizing and Benchmarking Prompt Injection Attacks and Defenses},
  author={Liu, Yupei and Jia, Yuqi and Geng, Runpeng and Jia, Jinyuan and Gong, Neil Zhenqiang},
  booktitle={33rd USENIX Security Symposium (USENIX Security 24)},
  pages={1831--1847},
  year={2024},
  month={August},
  publisher={USENIX Association},
  address={Philadelphia, PA},
  isbn={978-1-939133-44-1},
  url={https://www.usenix.org/conference/usenixsecurity24/presentation/liu-yupei}
}

@inproceedings{jiang2026agentlab,
  title={{AgentLAB}: Benchmarking {LLM} Agents against Long-Horizon Attacks},
  author={Jiang, Tanqiu and Wang, Yuhui and Liang, Jiacheng and Wang, Ting},
  booktitle={Proceedings of the 43rd International Conference on Machine Learning (ICML 2026)},
  year={2026},
  publisher={Proceedings of Machine Learning Research},
  url={https://icml.cc/virtual/2026/poster/65640}
}

@misc{pulipaka2026hiddeninmemory,
  title={Hidden in Memory: Sleeper Memory Poisoning in {LLM} Agents},
  author={Pulipaka, Sidharth and Hlebik, Stanislau and Raghav, Leonidas and Abdelnabi, Sahar and Raina, Vyas and Sheth, Ivaxi and Fritz, Mario},
  howpublished={arXiv preprint arXiv:2605.15338},
  year={2026},
  url={https://arxiv.org/abs/2605.15338}
}

@misc{zhao2026clawguard,
  title={{ClawGuard}: A Runtime Security Framework for Tool-Augmented {LLM} Agents Against Indirect Prompt Injection},
  author={Zhao, Wei and Li, Zhe and Zhang, Peixin and Sun, Jun},
  howpublished={arXiv preprint arXiv:2604.11790},
  year={2026},
  url={https://arxiv.org/abs/2604.11790}
}

@inproceedings{chen2024secalign,
  title={{SecAlign}: Defending Against Prompt Injection with Preference Optimization},
  author={Chen, Sizhe and Zharmagambetov, Arman and Mahloujifar, Saeed and Chaudhuri, Kamalika and Wagner, David and Guo, Chuan},
  booktitle={Proceedings of the 2025 ACM SIGSAC Conference on Computer and Communications Security},
  pages={2833--2847},
  year={2025},
  publisher={Association for Computing Machinery},
  address={New York, NY, USA},
  doi={10.1145/3719027.3744836},
  url={https://doi.org/10.1145/3719027.3744836}
}

@inproceedings{chen2024struq,
  title={{StruQ}: Defending Against Prompt Injection with Structured Queries},
  author={Chen, Sizhe and Piet, Julien and Sitawarin, Chawin and Wagner, David},
  booktitle={34th USENIX Security Symposium (USENIX Security 25)},
  pages={2383--2400},
  year={2025},
  month={August},
  address={Seattle, WA},
  publisher={USENIX Association},
  isbn={978-1-939133-52-6},
  url={https://www.usenix.org/conference/usenixsecurity25/presentation/chen-sizhe}
}

@misc{wang2026agentwatcher,
  title={{AgentWatcher}: A Rule-based Prompt Injection Monitor},
  author={Wang, Yanting and Zou, Wei and Geng, Runpeng and Jia, Jinyuan},
  howpublished={arXiv preprint arXiv:2604.01194},
  year={2026},
  url={https://arxiv.org/abs/2604.01194}
}

@misc{costa2025fides,
  title={Securing {AI} Agents with Information-Flow Control},
  author={Costa, Manuel and K{\"o}pf, Boris and Kolluri, Aashish and Paverd, Andrew and Russinovich, Mark and Salem, Ahmed and Tople, Shruti and Wutschitz, Lukas and Zanella-B{\'e}guelin, Santiago},
  howpublished={arXiv preprint arXiv:2505.23643},
  year={2025},
  url={https://arxiv.org/abs/2505.23643}
}

@misc{cai2026neurotaint,
  title={Ghost in the Agent: Redefining Information Flow Tracking for {LLM} Agents},
  author={Cai, Yuandao and Tang, Wensheng and Wen, Cheng and Qin, Shengchao},
  howpublished={arXiv preprint arXiv:2604.23374},
  year={2026},
  url={https://arxiv.org/abs/2604.23374}
}

@article{siddiqui2024permissiveifa,
  title={Permissive Information-Flow Analysis for Large Language Models},
  author={Siddiqui, Shoaib Ahmed and Gaonkar, Radhika and K{\"o}pf, Boris and Krueger, David and Paverd, Andrew and Salem, Ahmed and Tople, Shruti and Wutschitz, Lukas and Xia, Menglin and Zanella-B{\'e}guelin, Santiago},
  journal={Transactions on Machine Learning Research},
  year={2025},
  month={October},
  issn={2835-8856},
  url={https://openreview.net/forum?id=ufYRO8y3mr}
}

@misc{groth2013provoverview,
  title={{PROV}-Overview: An Overview of the {PROV} Family of Documents},
  author={Groth, Paul and Moreau, Luc},
  year={2013},
  howpublished={W3C Working Group Note},
  url={https://www.w3.org/TR/prov-overview/}
}

@inproceedings{rottger2023xstest,
  title={{XSTest}: A Test Suite for Identifying Exaggerated Safety Behaviours in Large Language Models},
  author={R{\"o}ttger, Paul and Kirk, Hannah and Vidgen, Bertie and Attanasio, Giuseppe and Bianchi, Federico and Hovy, Dirk},
  booktitle={Proceedings of the 2024 Conference of the North American Chapter of the Association for Computational Linguistics: Human Language Technologies (Volume 1: Long Papers)},
  pages={5377--5400},
  year={2024},
  month={June},
  address={Mexico City, Mexico},
  publisher={Association for Computational Linguistics},
  doi={10.18653/v1/2024.naacl-long.301},
  url={https://aclanthology.org/2024.naacl-long.301/}
}

@inproceedings{jia2024taskshield,
  title={The Task Shield: Enforcing Task Alignment to Defend Against Indirect Prompt Injection in {LLM} Agents},
  author={Jia, Feiran and Wu, Tong and Qin, Xin and Squicciarini, Anna},
  booktitle={Proceedings of the 63rd Annual Meeting of the Association for Computational Linguistics (Volume 1: Long Papers)},
  pages={29680--29697},
  year={2025},
  month={July},
  address={Vienna, Austria},
  publisher={Association for Computational Linguistics},
  doi={10.18653/v1/2025.acl-long.1435},
  url={https://aclanthology.org/2025.acl-long.1435/}
}

@misc{shi2025progent,
  title={{Progent}: Securing {AI} Agents with Privilege Control},
  author={Shi, Tianneng and He, Jingxuan and Wang, Zhun and Li, Hongwei and Wu, Linyu and Guo, Wenbo and Song, Dawn},
  howpublished={arXiv preprint arXiv:2504.11703},
  year={2025},
  url={https://arxiv.org/abs/2504.11703}
}

@misc{wang2026authgraph,
  title={Aligning Provenance with Authorization: A Dual-Graph Defense for {LLM} Agents},
  author={Wang, Peiran and Li, Ying and Tian, Yuan},
  howpublished={arXiv preprint arXiv:2605.26497},
  year={2026},
  url={https://arxiv.org/abs/2605.26497}
}

@misc{kang2025intentguard,
  title={Mitigating Indirect Prompt Injection via Instruction-Following Intent Analysis},
  author={Kang, Mintong and Xiang, Chong and Kariyappa, Sanjay and Xiao, Chaowei and Li, Bo and Suh, Edward},
  howpublished={arXiv preprint arXiv:2512.00966},
  year={2025},
  url={https://arxiv.org/abs/2512.00966}
}

@misc{debenedetti2025camel,
  title={Defeating Prompt Injections by Design},
  author={Debenedetti, Edoardo and Shumailov, Ilia and Fan, Tianqi and Hayes, Jamie and Carlini, Nicholas and Fabian, Daniel and Kern, Christoph and Shi, Chongyang and Terzis, Andreas and Tram{\`e}r, Florian},
  howpublished={arXiv preprint arXiv:2503.18813},
  year={2025},
  url={https://arxiv.org/abs/2503.18813}
}

@inproceedings{he2026attriguard,
  title={{AttriGuard}: Defeating Indirect Prompt Injection in {LLM} Agents via Causal Attribution of Tool Invocations},
  author={He, Yu and Zhu, Haozhe and Li, Yiming and Shao, Shuo and Yao, Hongwei and Liu, Zhihao and Qin, Zhan},
  booktitle={35th USENIX Security Symposium (USENIX Security 26)},
  year={2026},
  month={August},
  address={Baltimore, MD},
  publisher={USENIX Association},
  note={To appear},
  url={https://www.usenix.org/conference/usenixsecurity26/presentation/he-yu}
}

@misc{ma2026autodojo,
  title={{AutoDojo}: Adaptive Black-Box Attacks Reveal the Limits of {IPI} Defenses and Task-Specification Effects in {LLM} Agents},
  author={Ma, Xinhang and Li, Taoran and Xiao, Chaowei and Yu, Zhiyuan and Zhang, Ning and Vorobeychik, Yevgeniy},
  howpublished={arXiv preprint arXiv:2606.15057},
  year={2026},
  url={https://arxiv.org/abs/2606.15057}
}

@inproceedings{zhan2025adaptive,
  title={Adaptive Attacks Break Defenses Against Indirect Prompt Injection Attacks on {LLM} Agents},
  author={Zhan, Qiusi and Fang, Richard and Panchal, Henil Shalin and Kang, Daniel},
  booktitle={Findings of the Association for Computational Linguistics: NAACL 2025},
  pages={7116--7132},
  year={2025},
  address={Albuquerque, New Mexico},
  publisher={Association for Computational Linguistics},
  doi={10.18653/v1/2025.findings-naacl.395},
  url={https://aclanthology.org/2025.findings-naacl.395/}
}

@inproceedings{yin2026pismith,
  title={{PISmith}: Reinforcement Learning-based Red Teaming for Prompt Injection Defenses},
  author={Yin, Chenlong and Geng, Runpeng and Wang, Yanting and Jia, Jinyuan},
  booktitle={Proceedings of the 3rd Conference on Language Modeling (COLM 2026)},
  year={2026},
  note={To appear},
  url={https://colmweb.org/AcceptedPapers.html}
}

@misc{he2026reta,
  title={Defending against Adaptive Prompt Injection Attacks via Reasoning-enabled Task Alignment},
  author={He, Lipeng and Wang, Yihan and Zhang, Jiawen and Asokan, N.},
  howpublished={arXiv preprint arXiv:2606.15441},
  year={2026},
  url={https://arxiv.org/abs/2606.15441}
}

@misc{li2026agentdyn,
  title={{AgentDyn}: Are Your Agent Security Defenses Deployable in Real-World Dynamic Environments?},
  author={Li, Hao and Wen, Ruoyao and Shi, Shanghao and Zhang, Ning and Vorobeychik, Yevgeniy and Xiao, Chaowei},
  howpublished={arXiv preprint arXiv:2602.03117},
  year={2026},
  url={https://arxiv.org/abs/2602.03117}
}

@misc{jiang2023mistral,
  title={{Mistral 7B}},
  author={Jiang, Albert Q. and Sablayrolles, Alexandre and Mensch, Arthur and Bamford, Chris and Chaplot, Devendra Singh and {de las Casas}, Diego and Bressand, Florian and Lengyel, Gianna and Lample, Guillaume and Saulnier, Lucile and Lavaud, L{\'e}lio Renard and Lachaux, Marie-Anne and Stock, Pierre and Le Scao, Teven and Lavril, Thibaut and Wang, Thomas and Lacroix, Timoth{\'e}e and El Sayed, William},
  howpublished={arXiv preprint arXiv:2310.06825},
  year={2023},
  doi={10.48550/arXiv.2310.06825},
  url={https://arxiv.org/abs/2310.06825}
}

@misc{mistralv03modelcard,
  title={{Mistral-7B-Instruct-v0.3} Model Card},
  organization={Mistral AI},
  howpublished={Hugging Face model card},
  year={2024},
  url={https://huggingface.co/mistralai/Mistral-7B-Instruct-v0.3},
  note={Accessed 2026-08-02}
}

\end{document}